\pdfoutput=1
\documentclass[prl,aps,twocolumn,showpacs,preprintnumbers,amsmath,amssymb,superscriptaddress,floatfix]{revtex4-1}
\usepackage{graphicx}
\usepackage{amsmath} 
\usepackage{bm}
\usepackage{natbib}
\usepackage{booktabs} 
\usepackage{color}
\begin{document}


\title{Ultrahigh Thermal Conductivity and Mirror-Parity Constraints in a Corrugated Two-Dimensional Oxide} 

\author{Babu Ram} 
\email{babuaphy07@gmail.com}
\affiliation{Department of Chemistry, Ulsan National Institute of Science and Technology (UNIST), Ulsan, 44919, Republic of Korea} 

\author{Arun S. Nissimagoudar}
\affiliation{OLA Battery Innovation Center, OLA Cell Technologies, Bangalore, 560100, India} 

\author{Weichao Wang}
\affiliation{College of Electronic Information and Optical Engineering, Nankai University, Tianjin 300350, China}

\author{Geunsik Lee}
\email{gslee@unist.ac.kr} 
\affiliation{Department of Chemistry, Ulsan National Institute of Science and Technology (UNIST), Ulsan, 44919, Republic of Korea}

\begin{abstract}Contrary to the expectation of poor heat transport in oxides, first-principles calculations predict an ultrahigh lattice thermal conductivity of 756 $\mathrm{W\,m^{-1}\,K^{-1}}$ at 300 K in corrugated two-dimensional B$_2$O$_2$. Acoustic phonons dominate heat transport, with the flexural acoustic (ZA) branch contributing 40.62$\%$. Triplet resolved analysis including acoustic and optical partners finds that mirror forbidden three phonon scattering is negligible within numerical precision for sampled ZA modes. These results highlight horizontal mirror symmetry as a constraint on phonon scattering in corrugated oxides. 
\end{abstract} 
\maketitle



\noindent\textit{Introduction—}Efficient thermal management is increasingly important as electronic and optoelectronic devices become smaller and operate at higher power densities, motivating the search for materials with exceptional heat transport and a clear understanding of the phonon mechanisms that govern it. Two dimensional materials provide a particularly useful platform for exploring such transport phenomena. Suspended graphene exhibits an extraordinary room temperature thermal conductivity of 3000--5000~$\mathrm{W\,m^{-1}\,K^{-1}}$~\cite{Bala2008,Gosh2008,Niki2009}, while monolayer $h$-BN reaches 600--1134~$\mathrm{W\,m^{-1}\,K^{-1}}$~\cite{Lin2011,Qi2019,Fan2019}. Their exceptional heat transport is closely associated with horizontal mirror selection rules that restrict three phonon scattering and support long phonon lifetimes. These constraints can survive finite layer thickness, as demonstrated in mirror symmetric D-AA diamane~\cite{Zhu2019}. Representative two dimensional oxides, however, exhibit substantially lower thermal conductivities: BeO (385 W\,m$^{-1}$\,K$^{-1}$)~\cite{Zheng2015,Zha2021,Mor2021}, MgO (64 W\,m$^{-1}$\,K$^{-1}$)~\cite{Zheng2015,Mor2021}, CaO (15 W\,m$^{-1}$\,K$^{-1}$)~\cite{Zheng2015,Mor2021}, and ZnO (4.5 W\,m$^{-1}$\,K$^{-1}$)~\cite{Wan2017}. This contrast motivates the search for light element oxide networks that combine stiff bonding with symmetry constraints on phonon scattering.\\ 
\begin{figure*}[ht]
\centering 
\includegraphics[width=0.8\textwidth]{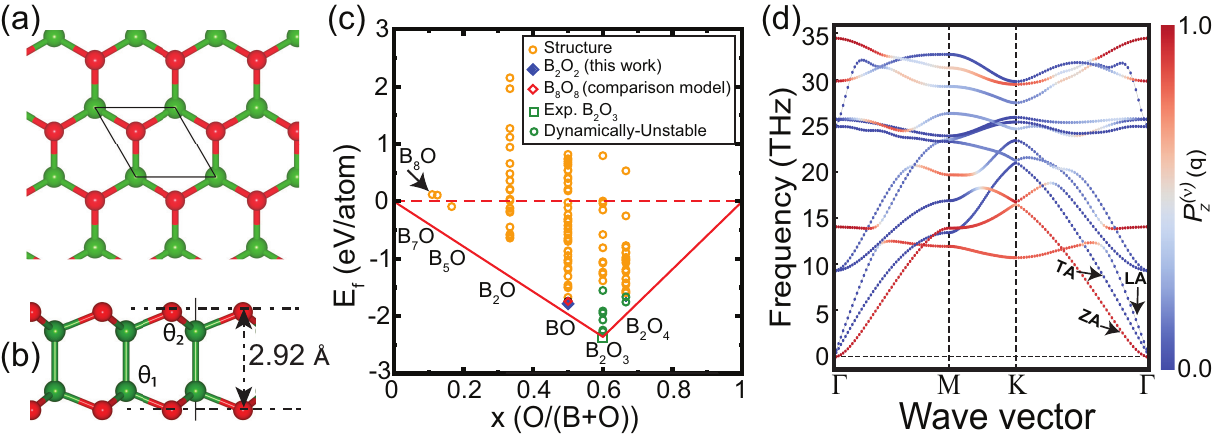} 
\caption{Structure, energetic stability, and lattice dynamics of monolayer B$_2$O$_2$. (a,b) Top and side views of the optimized $P\bar{6}m2$ structure with a physical thickness of 2.92 \AA{}. 
(c) Formation energies of the considered two dimensional boron oxides versus oxygen fraction, $x=\mathrm{O}/(\mathrm{B}+\mathrm{O})$. The blue diamond denotes corrugated B$_2$O$_2$, which lies approximately $0.18~\mathrm{eV/atom}$ above the convex hull of the included structures. The B$_8$O$_8$ sheet (red diamond) is included as a structural comparison model~\cite{Cla2010,Per2023,Boh2023}, while the B$_2$O$_3$ reference (green square) corresponds to the monolayer synthesized on Pt(111)~\cite{Ferlat2012,T2025}. Green open circles indicate dynamically unstable structures (Fig.~S1 of the Supplemental Material). Solid red lines denote the convex hull, and the dashed red line marks zero formation energy. The B$_5$O, B$_7$O, and B$_8$O structures are from Ref.~\cite{Zh2017}. (d) Phonon dispersion color coded by the normalized out-of-plane eigenvector projection $P_z^{(\nu)}(\mathbf q)$, from predominantly in-plane (blue) to out-of-plane (red) polarization (Sec. S2.1 and Table S3 of the Supplemental Material).}
\label{Fig:1} 
\end{figure*}
Boron oxides provide a promising platform because they can sustain extended B-O-B networks. Monolayer B$_2$O$_3$ has been synthesized on Pt(111)~\cite{T2025}, demonstrating the experimental accessibility of two dimensional boron oxide networks. A recent high throughput screening study also identified monolayer B$_2$O$_2$ as a promising thermal transport candidate and predicted a room temperature lattice thermal conductivity of 598 $\mathrm{W\,m^{-1}\,K^{-1}}$~\cite{Chen2025}. Building on this prediction, we examine the thermodynamic stability and bonding characteristics of monolayer B$_2$O$_2$, which remain insufficiently established. We investigate how horizontal mirror symmetry constrains phonon scattering in this corrugated structure, where modes can acquire mixed Cartesian polarization. We also distinguish the contribution of optical phonons to heat transport from their participation as scattering partners of acoustic modes.\\ 
 \indent Here, we independently identify the monolayer through a particle swarm optimization (PSO) structure search using CALYPSO~\cite{Wang2012} and assess its structural stability and bonding through first-principles calculations with VASP~\cite{Kress1996,KressPRB1996,KresseJ1999}. An iterative solution of the phonon Boltzmann transport equation using ShengBTE~\cite{Sheng2014} predicts a lattice thermal conductivity of 756 $\mathrm{W\,m^{-1}\,K^{-1}}$ at 300 K. Our mode resolved analysis distinguishes mirror parity from Cartesian polarization by explicitly accounting for the permutation of symmetry related atoms and the transformation of their displacements. We then identify the acoustic and optical partners contributing to symmetry allowed ZA three phonon scattering. We show that optical phonons contribute only modestly to heat transport, yet channels involving optical partners dominate ZA scattering at M, revealing their distinct roles as heat carriers and scattering partners.\\
\textit{Results and discussion—}Fig.~\ref{Fig:1} summarizes the structure, thermodynamic stability, and phonon band structure of monolayer B$_2$O$_2$. The optimized top and side views (Figs.~\ref{Fig:1}(a) and \ref{Fig:1}(b)) reveal a corrugated hexagonal network belonging to the $P\bar{6}m2$ space group, with an in-plane lattice parameter of $2.44$ \AA{} and a physical thickness of $2.92$ \AA{}. The corresponding bond lengths and angles are listed in Table S1 of the Supplemental Material. The thermodynamic stability of B$_2$O$_2$ is assessed from the formation energy ($E_f$) defined in Sec. S1.2 of the Supplemental Material, with the corresponding results shown in Fig.~\ref{Fig:1}(c). B$_2$O$_2$ lies $\sim$ 0.18 $\mathrm{eV/atom}$ above the convex hull of the considered two dimensional B-O structures, indicating that it is metastable. At the same reduced B:O composition, B$_2$O$_2$ has a lower calculated energy than the B$_8$O$_8$ sheet~\cite{Cla2010,Per2023,Boh2023} included as a structural comparison model. By contrast, B$_2$O$_3$ forms a vertex of the calculated convex hull and has been synthesized as a monolayer on Pt(111)~\cite{T2025}. The optimized configurations, formation energies, and phonon dispersions of the low energy B-O structures marked by the green symbols in Fig.~\ref{Fig:1}(c) are provided in Fig.~S1, while the computational procedures underlying the structural, energetic, and phonon analyses are detailed in Secs. S1.1, S1.2, and S2 of the Supplemental Material.\\
 \indent The phonon dispersion in Fig.~\ref{Fig:1}(d) shows no significant imaginary frequencies apart from a small ZA softening near $\Gamma$, which is attributed to the numerical sensitivity of long wavelength flexural phonons. Convergence of the phonon dispersion with respect to the harmonic supercell size is demonstrated in Figs. S2(a,b). Consistent with this robust flexural dispersion, the long wavelength ZA branch yields a bending rigidity of approximately $30$ eV (Sec. S1.3, Fig. S2(c), and Table S2), reflecting the strong resistance of the $2.92$ \AA{} thick corrugated B$_2$O$_2$ layer to out-of-plane deformation and its strongly bonded B--O/B--B framework. Although the quadratic ZA mode has $v_g\rightarrow0$ as $\mathbf q\rightarrow0$, its group velocity increases away from $\Gamma$, enabling finite wave vector flexural phonons to participate effectively in heat transport. This out-of-plane rigidity is complemented by the strong in-plane mechanical response: linear fits to the stress-strain curves (Fig.~S3(a)) yield uniaxial Young's moduli of $558$ and $564$ GPa along the two in-plane directions, respectively, together with an equibiaxial modulus of $680$ GPa (Sec. S1.4). The long wavelength LA and TA branches have sound velocities of $14.8$ and $9.2$ $\mathrm{km\,s^{-1}}$, respectively.\\
 \begin{figure*}[ht]  
\centering 
\includegraphics[width=0.8\textwidth]{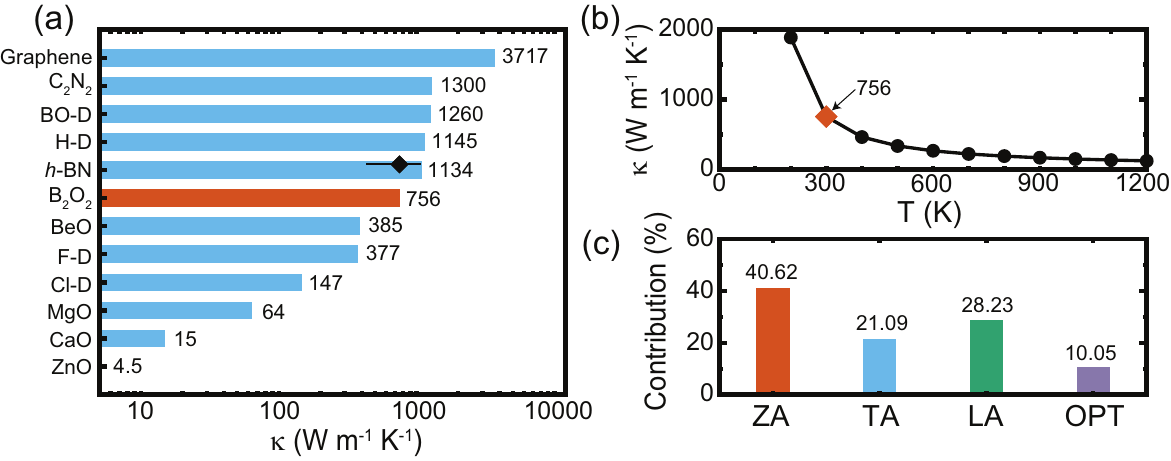} 
\caption{Lattice thermal transport in monolayer B$_2$O$_2$. (a) Lattice thermal conductivity $\kappa$ at 300 K compared with representative two dimensional materials: graphene~\cite{Pen2016}, diamanes (H, F, Cl, and BO)~\cite{Mos2020,Ram2026}, C$_{2}$N$_{2}$~\cite{Chen2025}, $h$-BN~\cite{Fan2019}, and oxides~\cite{Mor2021}. B$_2$O$_2$ is highlighted in orange; the horizontal axis is logarithmic. The black diamond and horizontal error bar indicate the experimental $h$-BN value, 751$\pm$340 $\mathrm{W\,m^{-1}\,K^{-1}}$~\cite{Qi2019}. D denotes diamane. (b) Temperature dependence of $\kappa$ for B$_2$O$_2$; the orange diamond marks $\kappa$ = 756 $\mathrm{W\,m^{-1}\,K^{-1}}$ at 300 K. (c) Contributions to $\kappa$ at 300 K from the out-of-plane transverse acoustic (ZA), in-plane transverse acoustic (TA), longitudinal acoustic (LA), and optical (OPT) phonon branches.}
\label{Fig:2} 
\end{figure*} 
 \indent The mechanical robustness is further supported by \textit{ab initio} molecular dynamics simulations, which show that the B$_2$O$_2$ framework remains intact without appreciable reconstruction up to $1000$ K (Figs. S3(b)-S3(d)). The mechanical and finite temperature robustness of B$_2$O$_2$ is further rationalized by its bonding characteristics, as revealed by the electron localization function (ELF) in Figs. S4(a) and S4(b). The ELF shows greater electron localization toward O along the B--O bonds, whereas the electron distribution between neighboring B atoms is approximately symmetric. Consistently, Bader analysis indicates an average charge transfer of $\sim$ 1.54$e$ per B atom to O, supporting the coexistence of polar covalent B--O and covalent B--B bonding. The electronic structure further supports this picture, with a predominantly O-$p_z$ valence band edge accompanied by a B-$p_z$ contribution, indicating appreciable B-O orbital hybridization. The band structure and projected density of states (Sec. S1.5 and Fig. S4(c) of the Supplemental Material) further reveal an indirect PBE band gap of 4.74 eV, supporting the neglect of electronic heat transport under intrinsic conditions.\\ 
\begin{figure}[!ht] 
\centering
\includegraphics[width=0.7\columnwidth]{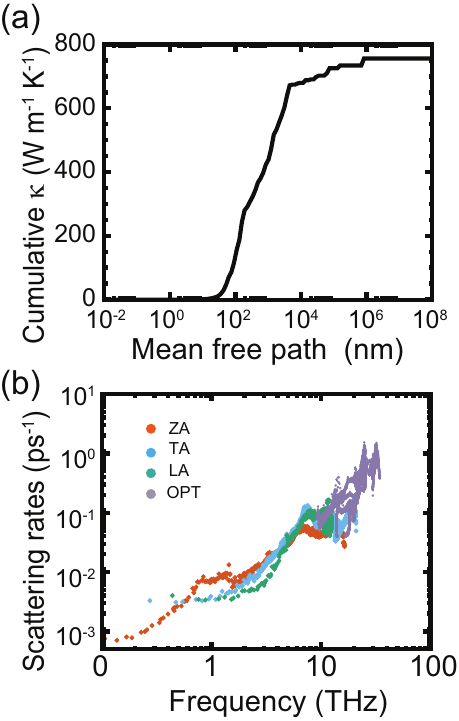} 
\caption{(a) Cumulative lattice thermal conductivity, ($\kappa$), as a function of phonon mean free path at 300 K. (b) Mode resolved three phonon scattering rates as a function of frequency for the flexural acoustic (ZA), in-plane transverse acoustic (TA), longitudinal acoustic (LA), and optical (OPT) branches.}  
\label{Fig:3}
\end{figure} 
 \indent Near $\Gamma$, ZA is predominantly out-of-plane polarized, whereas TA and LA are predominantly in-plane polarized (Fig. 1(d), Sec. S2.1, and Table S3). Several branches acquire mixed Cartesian polarization at finite wave vector. Such mixing is compatible with definite mirror parity because $\sigma_h$ acts on both the displacement components and the symmetry related atoms. At $\Gamma$, the acoustic representation $\Gamma_{\rm ac}=A_2''(\mathrm{ZA})\oplus E'(\mathrm{TA,LA})$ identifies ZA as mirror odd and TA/LA as mirror even. The complete $\Gamma$-point representations and polarization projections are listed in Table S3; the explicitly evaluated finite wave vector mirror eigenvalues are reported in Sec. S2.2 and Table S4.\\
 \indent We next connect this symmetry classification to thermal transport and phonon scattering. The iterative BTE solution predicts $\kappa$ = 756 $\mathrm{W\,m^{-1}\,K^{-1}}$ at 300 K using an effective thickness of $5.88$ \AA{} derived from the optimized layered reference structure (Sec. S3.3). Under the respective reported thickness conventions, this value is nearly twice that of BeO (385 $\mathrm{W\,m^{-1}\,K^{-1}}$)~\cite{Mor2021}, comparable to the experimental monolayer $h$-BN value of 751 $\pm$ 340 $\mathrm{W\,m^{-1}\,K^{-1}}$~\cite{Qi2019}, and below the theoretical $h$-BN estimate of 1134 $\mathrm{W\,m^{-1}\,K^{-1}}$~\cite{Fan2019} (Fig.~\ref{Fig:2}(a)). Among the oxides compared in Fig.~S5(a,b), B$_2$O$_2$ has an LA velocity comparable to BeO and a higher TA velocity. Its position above both fitted velocity conductivity trends further motivates a microscopic examination of its phonon transport. The conductivity decreases monotonically with temperature, consistent with increasing anharmonic scattering (Fig.~\ref{Fig:2}(b)). At 300 K, ZA, TA, LA, and optical branches contribute $40.62\%$, $21.09\%$, $28.23\%$, and $10.05\%$, respectively (Fig.~\ref{Fig:2}(c)). Acoustic branches therefore carry nearly $90\%$ of the heat, with ZA providing the largest individual contribution. Figure~\ref{Fig:3}(a) shows the cumulative lattice thermal conductivity as a function of phonon mean free path ($\Lambda$). The conductivity weighted median mean free path ($\Lambda_{50}$), at which the cumulative conductivity reaches half its total value, is approximately $0.57~\mu\mathrm{m}$. The accumulation reaches $90\%$ at $\Lambda\simeq12~\mu\mathrm{m}$, indicating substantial heat transport by phonons with micrometer scale mean free paths. The normalized cumulative conductivity, obtained by dividing by the total conductivity, is shown in Fig. S6(a). Low frequency acoustic modes generally exhibit smaller three phonon scattering rates than most optical modes (Fig.~\ref{Fig:3}(b)), consistent with their lifetime distribution in (Fig. S6(b)). The large negative Gr\"uneisen parameters of low-frequency modes indicate strong strain sensitivity (Fig.~S6(c)), but their magnitudes alone do not determine the phonon scattering rates. We therefore examine the mirror-parity selection rules governing three-phonon scattering of ZA modes.\\ 
\begin{figure*}[!ht] 
\centering
\includegraphics[width=0.9\textwidth]{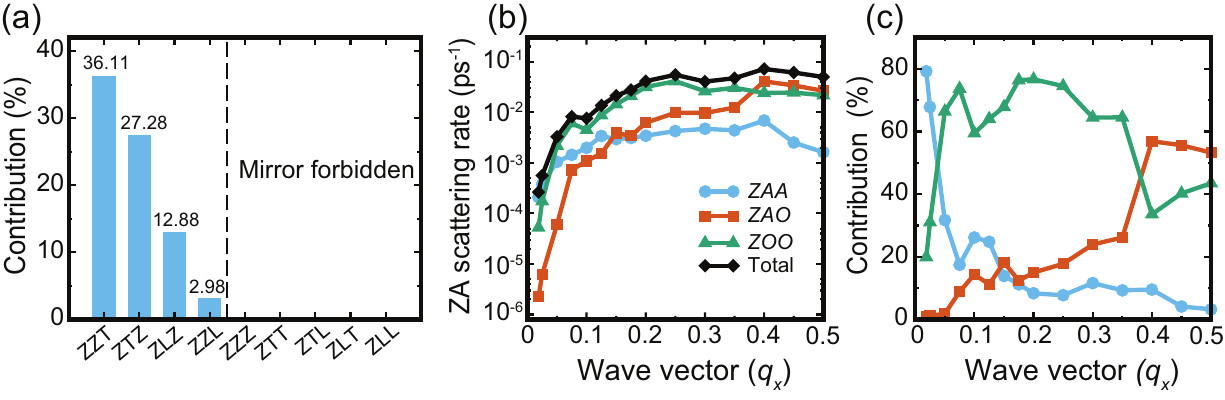}
\caption{Channel resolved three phonon scattering of ZA phonons in monolayer B$_2$O$_2$ at 300 K. (a) Contributions of the nine ordered acoustic partner channels to the total three phonon scattering rate of the ZA mode at $q$ = (0.01875, 0, 0). The four mirror allowed channels contribute 79.25$\%$ of the total rate, whereas the five mirror forbidden channels are numerical residuals and vanish on the plotted scale. (b) ZA scattering rates along $\Gamma$-M, resolved into acoustic-acoustic ($ZAA$), acoustic optical ($ZAO$), and optical-optical ($ZOO$) partner sectors, together with their total. (c) Contributions of the three sectors expressed as percentages of the complete scattering rate of each initial ZA mode. In each channel label, the first Z denotes the initial ZA mode and the last two symbols identify the partner branches. $Z$, $T$, and $L$ denote the continuously tracked ZA, TA, and LA branches; $A\in\{Z,T,L\}$, and $O$ denotes any of the nine optical branches. $ZAO$ includes both partner orderings. Wave vectors $\mathbf{q}=(q_x,0,0)$ are given in reciprocal lattice coordinates, with M at $q_{x}$ = 0.5.}  
\label{Fig:4} 
\end{figure*}
 \indent For in-plane phonon wave vectors, horizontal reflection $\sigma_h$ leaves $\mathbf q$ invariant. Let $m_\lambda=\pm1$ denote the mirror eigenvalue of phonon mode $\lambda$. The complete mirror operation includes the permutation of symmetry related atoms, the Cartesian transformation of their displacements, and the appropriate Bloch phase. The three phonon matrix element transforms as 

\begin{equation}
\Phi_{\lambda\lambda'\lambda''}^{\pm}
\xrightarrow{\sigma_h}
m_\lambda m_{\lambda'}m_{\lambda''}
\Phi_{\lambda\lambda'\lambda''}^{\pm} 
\label{eq:mirror_transformation}
\end{equation}
where $+$ and $-$ denote absorption and emission, respectively. Mirror invariance therefore requires
\begin{equation}
\left(1-m_\lambda m_{\lambda'}m_{\lambda''}\right)
\Phi_{\lambda\lambda'\lambda''}^{\pm}=0 
\label{eq:mirror_invariance}
\end{equation}
A nonzero matrix element thus requires $m_\lambda m_{\lambda'}m_{\lambda''}=+1$; triplets with parity product $-1$ are forbidden in the exact $\sigma_h$-symmetric limit. To maintain consistent acoustic branch labels at finite wave vector without relying on local frequency ordering, we continuously track the three acoustic branches over the $160\times160\times1$ mesh using maximum eigenvector overlap matching. The tracked branches are denoted by $Z$, $T$, and $L$, while $O$ collectively denotes the remaining nine optical modes. Branch tracking establishes mode identity; mirror parity is evaluated separately using the complete symmetry operation for every participating acoustic and optical mode.\\ 
 \indent Fig.~\ref{Fig:4}(a) resolves the nine ordered acoustic partner channels of a representative ZA mode at $\mathbf q=(0.01875,0,0)$ and T = 300 K. Their contributions are normalized to the complete three phonon scattering rate of that mode. The four mirror allowed channels, $ZZT$, $ZZL$, $ZTZ$, and $ZLZ$, contribute $79.25\%$ in total, whereas the five forbidden channels, $ZZZ$, $ZTT$, $ZTL$, $ZLT$, and $ZLL$, are negligible within numerical precision. The remaining contributions arise from acoustic-optical ($ZAO$, $0.85\%$) and optical-optical ($ZOO$, $19.90\%$) partner sectors. Fig.~\ref{Fig:4}(b,c) show the absolute rates and relative contributions of the $ZAA$, $ZAO$, and $ZOO$ sectors along $\Gamma$-M, where $A$ denotes any tracked acoustic branch. The $ZAO$ sector includes both acoustic-optical partner orderings. Optical containing channels account for $20.75\%$ of the rate at $q_x=0.01875$ and $96.76\%$ at M ($q_x=0.50$). The largest sampled total rate occurs at $q_x=0.40$, where these channels contribute $90.47\%$. Optical containing channels remain subject to the same mirror selection rule. For an initial ZA mode with $m_{\lambda}=-1$, an allowed triplet requires $m_{\lambda'}m_{\lambda''}=-1$: the two partner modes must have opposite mirror parities, including when one or both are optical.\\ 
 \indent The mirror forbidden contribution to the ZA scattering rate remains negligible within numerical precision throughout the sampled $\Gamma$-M branch (Sec. S4.1 and Tables S5 and S6). Optical containing channels account for nearly $97\%$ of the ZA scattering rate at M, although optical phonons carry only $10.05\%$ of the heat at 300 K, underscoring their distinct roles as heat carriers and scattering partners.\\
 \indent We additionally separate the acoustic three phonon scattering into Normal (N) and Umklapp (U) contributions. Normal processes dominate over the sampled near-$\Gamma$ region along $\Gamma$-M at 300 K, with the Umklapp fraction remaining below approximately $3\%$ (Sec. S4.2 and Table S7). This decomposition characterizes momentum conservation in the sampled processes and complements the mirror parity classification.\\ 
 \indent To explore possible routes toward experimental realization, we examine the interactions of B$_2$O$_2$ with Cu(111) and Ni(111) as candidate growth substrates. The calculated binding energy magnitudes of approximately 0.12--0.16 eV and equilibrium interface separations of 3.2--3.3 \AA{} indicate weak coupling of B$_2$O$_2$ to Cu(111) and Ni(111), accompanied by only negligible interfacial charge transfer of about 0.01$e$ (Sec. S5.1 and Figs. (S7,S8), and Table S8 of the Supplemental Material). These weak interactions suggest that both surfaces may serve as potential growth templates for B$_2$O$_2$; however, their effects on the vibrational stability and thermal transport of the supported structures require separate assessment.\\ 
\noindent\textit{Conclusions—}Corrugated monolayer B$_2$O$_2$ combines horizontal mirror symmetry with a room temperature lattice thermal conductivity of $756$ $\mathrm{W\,m^{-1}\,K^{-1}}$. Acoustic phonons carry approximately $90\%$ of the heat, with the ZA branch making the largest individual contribution. Explicit symmetry analysis shows that mixed Cartesian polarization is compatible with definite mirror parity, and mirror forbidden contributions to ZA three phonon scattering remain negligible within numerical precision at the sampled wave vectors along $\Gamma$-M, including the zone boundary. Optical phonons play a distinct scattering role: although they carry only about $10\%$ of the heat, optical containing channels account for nearly $97\%$ of the ZA scattering rate at M. These results show that mirror parity constraints persist even when optical partners dominate the symmetry-allowed ZA scattering. 
Weak calculated coupling to Cu(111) and Ni(111) motivates their consideration as candidate substrates for realizing B$_2$O$_2$, although substrate effects on its phonon spectrum and thermal transport remain to be established. Our findings show that structural corrugation can coexist with mirror selection rules that restrict phonon scattering, providing a symmetry based perspective on heat transport in corrugated two dimensional crystals.\\
\noindent\textit{Acknowledgement—}We acknowledge financial support from the Brain Pool Program (RS-2025-25440428), funded by the National Research Foundation of Korea (NRF). The DFT calculations were performed using computational resources provided by the Korea Institute of Science and Technology Information (KISTI).

\bibliography{Manuscript} 

\end{document}


\raggedbottom

\title{{\large Supplemental Material}\\[0.5em]
Ultrahigh Thermal Conductivity and Mirror-Parity Constraints in a Corrugated Two-Dimensional Oxide}

\author{Babu Ram}
\email{babuaphy07@gmail.com}
\affiliation{
Department of Chemistry,
Ulsan National Institute of Science and Technology (UNIST),
Ulsan 44919, Republic of Korea
}

\author{Arun S. Nissimagoudar}
\affiliation{
OLA Battery Innovation Center,
OLA Cell Technologies,
Bangalore 560100, India
}

\author{Weichao Wang}
\affiliation{College of Electronic Information and Optical Engineering, Nankai University, Tianjin 300350, China} 

\author{Geunsik Lee}
\email{gslee@unist.ac.kr}
\affiliation{
Department of Chemistry,
Ulsan National Institute of Science and Technology (UNIST),
Ulsan 44919, Republic of Korea
}

\maketitle

\thispagestyle{plain}
\vspace*{1.0cm}
\tableofcontents

%
%
%
%
%
%
%
%
%
%
%
%
%
%
%
\label{sec:interface}
\clearpage
\section{Computational details and structural stability}
\label{sec:computational}
\subsection{First-principles calculations and structure search}
\label{sec:dft}

First-principles calculations were performed within density functional theory (DFT), as implemented in the Vienna \textit{Ab initio} Simulation Package (VASP)~\cite{Kress1996,KressPRB1996}. The exchange correlation energy was treated using the Perdew-Burke-Ernzerhof generalized gradient approximation (PBE-GGA)~\cite{Perdew1996}, and the electron-ion interactions were described using the projector augmented wave (PAW) method~\cite{KresseJ1999,Blochl1994}. A plane wave kinetic energy cutoff of 600 eV was employed. The electronic energy was converged to $10^{-8}$ eV, and the atomic structures were relaxed until the Hellmann Feynman forces on every atom were smaller than $10^{-4}$ eV\,\AA$^{-1}$. The primitive cell of monolayer B$_2$O$_2$ contains four atoms. A vacuum region of at least 17 \AA{} was introduced along the out-of-plane direction to suppress interactions between periodically repeated monolayers. Brillouin zone integrations were performed using a $21\times21\times1$ Monkhorst Pack $\mathbf{k}$-point mesh~\cite{Monkhorst1976}. Long range dispersion interactions were included using Grimme's DFT-D3 correction~\cite{Grimme2010}.\\
 \indent Low energy boron oxide structures were explored using the CALYPSO particle swarm optimization method~\cite{WANG2012}, with searches performed for B$_2$O$_2$, B$_3$O$_3$, B$_4$O$_4$, B$_2$O$_3$, and B$_2$O$_4$. Additional starting structures for B$_5$O, B$_7$O, B$_8$O, B$_8$O$_8$, and B$_2$O$_3$ were adopted from previous theoretical studies. The B$_8$O$_8$ sheet was included for structural comparison, while the B$_2$O$_3$ reference corresponds to the monolayer synthesized on Pt(111) (Fig. 1(c) of the main text). All structures, including those adopted from the literature, were reoptimized, and their total energies and phonon dispersions were evaluated using the same DFT settings. 

\subsection{Energetic and dynamical stability}
\label{sec:stability}

To compare structures of different B-O stoichiometries on a common energetic basis and construct the corresponding convex hull, the formation energy per atom of each B$_n$O$_m$ structure was calculated as

\renewcommand{\thefigure}{S\arabic{figure}}
\begin{figure*}[t] 
\centering
\includegraphics[width=0.95\textwidth]{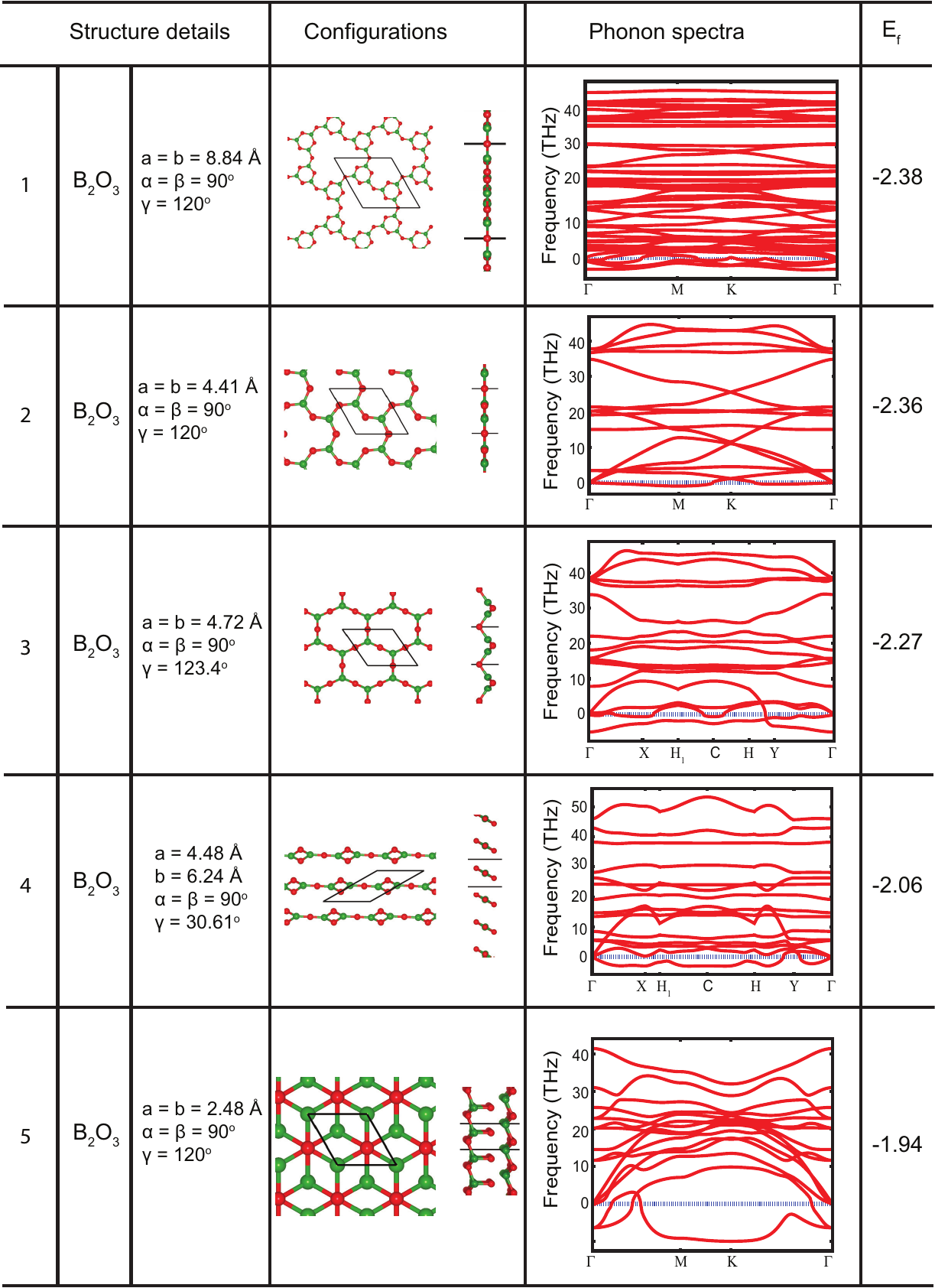}  
\end{figure*}

\renewcommand{\thefigure}{S\arabic{figure}} 
\begin{figure*}[t]
\centering
\includegraphics[width=0.9\textwidth]{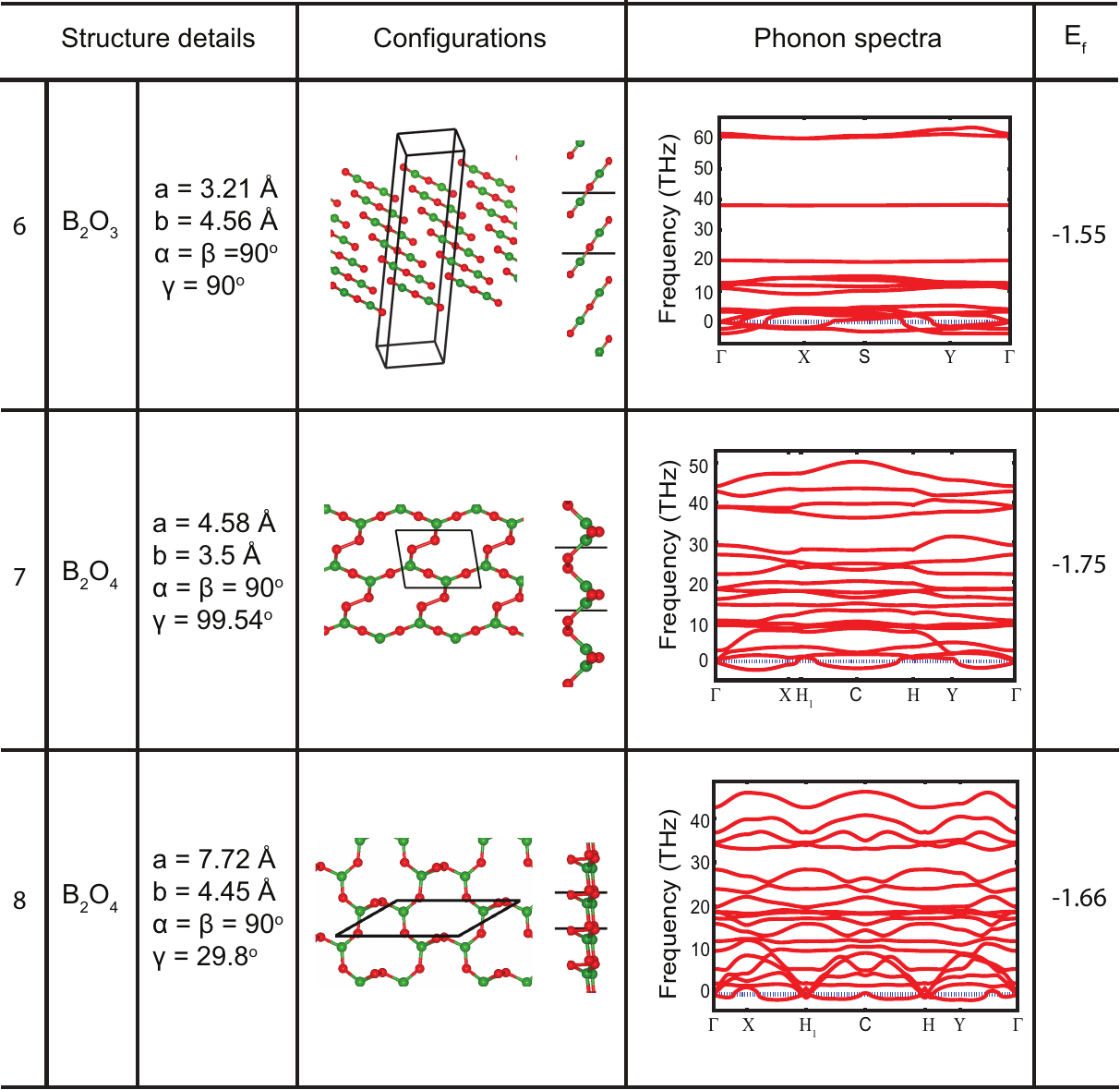} 
\caption{Optimized lattice parameters, top and side views of the atomic configurations, phonon dispersions, and formation energies ($E_{\mathrm f}$) of Structures 1 to 8, corresponding to the eight lowest energy configurations highlighted (green colored circles and square) in Fig. 1(c) of the main text. Imaginary phonon frequencies plotted below zero indicate dynamical instability. Structure (1) corresponds to the B$_2$O$_3$ monolayer previously predicted theoretically and subsequently realized experimentally on Pt(111)~\cite{Ferlat2012,T2025}; however, it exhibits significant dynamical instability in its free standing form, whereas B$_8$O$_8$ is found to be dynamically stable~\cite{Boh2023}.}   
\end{figure*} 

\begin{equation}
E_{\mathrm f}
=
\frac{
E_{\mathrm{B}_n\mathrm{O}_m}
-
nE_{\mathrm B}
-
mE_{\mathrm O} 
}{n+m} 
\label{eq:formation_energy}
\end{equation}
where $E_{\mathrm{B}_n\mathrm{O}_m}$ is the total energy of the fully optimized B$_n$O$_m$ unit cell, $E_{\mathrm B}=-6.81$ eV is the energy per atom of crystalline $\alpha$-boron, and $E_{\mathrm O}=\tfrac{1}{2}E_{\mathrm{O}_2}=-4.92$ eV is the energy per oxygen atom in an isolated O$_2$ molecule. Here, $n$ and $m$ denote the numbers of boron and oxygen atoms in the unit cell, respectively. Using these consistently calculated formation energies, the resulting convex hull analysis places B$_2$O$_2$ approximately $0.18$ eV\,atom$^{-1}$ above the convex hull of the B-O compositional space. 

The harmonic phonon dispersions calculated using $8\times8\times1$, $10\times10\times1$, and $12\times12\times1$ supercells nearly overlap (Fig. S2(b)), indicating convergence with respect to supercell size. The second order interatomic force constants obtained from the $12\times12\times1$ supercell were therefore adopted for the subsequent phonon, mirror parity, and thermal transport analyses; the corresponding dispersion is shown in Fig. S2(a).

A small near $\Gamma$ ZA softening remains, with an imaginary frequency magnitude of $\sim$ $0.02$ THz. The ZA scattering channel analysis in Sec. S4.1 begins at $\mathbf{q}=(0.01875,0,0)$ in reciprocal lattice coordinates, where the frequency is positive ($0.0630$ THz), and extends to the zone boundary. Other B-O configurations exhibiting appreciable imaginary phonon frequencies are marked by green symbols in 
Fig. 1(c) of the main text; their optimized structures and phonon dispersions are presented in Fig. S1.
\\
\\
\renewcommand{\thetable}{S\arabic{table}}
\begin{table*}[!ht]
\centering
\caption{Calculated lattice parameters, bond distances, physical thickness, and bond angles of monolayer B$_2$O$_2$.}
\label{tab:structural_properties}

\begin{ruledtabular} 
\begin{tabular}{cccc}

Lattice parameters &
Bond distances &
Thickness, $t$ &
Bond angles \\[2pt]

\hline

$a=b=2.44$ \AA{} &

\begin{tabular}{c}
$d_{\mathrm{B-B}}=1.77$ \AA{}\\
$d_{\mathrm{B-O}}=1.52$ \AA{}
\end{tabular}
&

$2.924$ \AA{}
&

\begin{tabular}{c}
$\theta_1=\angle\mathrm{BBO}=112.19^\circ$\\
$\theta_2=\angle\mathrm{BOB}=106.61^\circ$
\end{tabular}

\end{tabular}
\end{ruledtabular}
\end{table*}

\clearpage

\subsection{Bending rigidity}
\label{sec:bending}

The bending rigidity of monolayer B$_2$O$_2$ was obtained from the long wavelength flexural acoustic (ZA) dispersion. For a stress free two-dimensionalmembrane,
\begin{equation}
\omega_{\mathrm{ZA}}(q)=\sqrt{\frac{D}{\rho_{\mathrm{2D}}}}\,q^2
\equiv A q^2 
\end{equation}
where $D$ is the bending rigidity, $\rho_{\mathrm{2D}}$ is the areal mass density, and $A=\sqrt{D/\rho_{\mathrm{2D}}}$. The angular frequency is related to the phonon frequency plotted in THz by $\omega_{\mathrm{ZA}}=2\pi\Omega_{\mathrm{ZA}}$.  Fitting $\Omega_{\mathrm{ZA}}$ against $q^2$ in the selected long wavelength region (Fig. S2(c)) yields a bending rigidity  of $D=30.2$ eV. The corresponding group velocity is $v_g=\partial\omega_{\mathrm{ZA}}/\partial q=2Aq$, which vanishes as $q\rightarrow0$ and increases linearly with $q$ within the fitted regime.

\renewcommand{\thefigure}{S\arabic{figure}} 
\begin{figure*}[t]  
\centering
\includegraphics[width=0.7\textwidth]{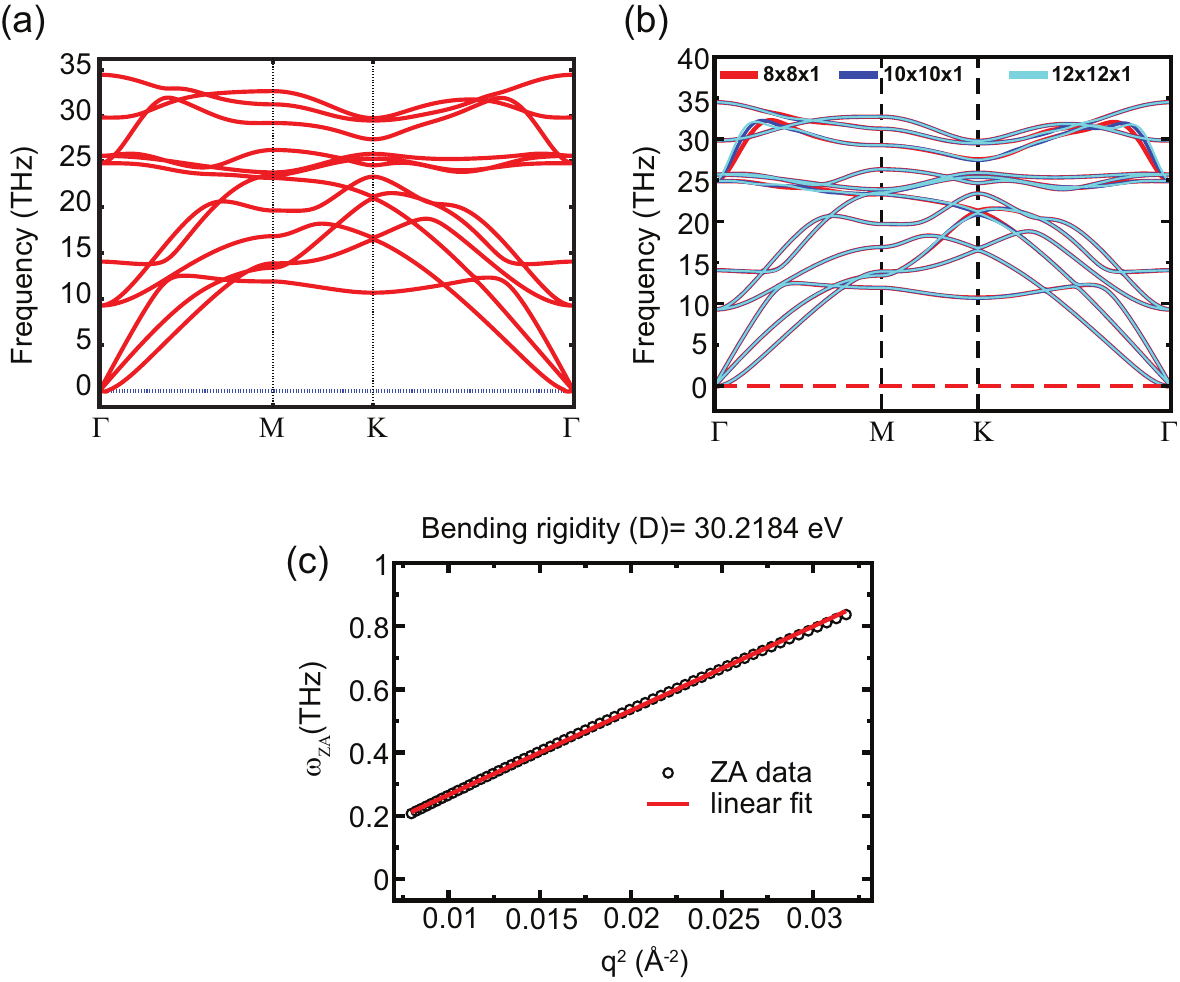}
\caption {(a) Phonon dispersion calculated using the second-order interatomic force constants obtained from a $12\times12\times1$ supercell. (b) Comparison of the phonon dispersions obtained using $8\times8\times1$, $10\times10\times1$, and $12\times12\times1$ second order force constant supercells. The horizontal dashed line denotes zero frequency. (c) Long wavelength ZA phonon frequency $\Omega_{\mathrm{ZA}}$ as a function of $q^2$. The linear fit yields a bending rigidity of $D=30.2$ eV.}
\end{figure*} 

\renewcommand{\thetable}{S\arabic{table}}
\begin{table}[H]
\caption{Comparison of the bending rigidities of representative two-dimensionalmaterials.}
\label{tab:bending_rigidity}

\centering
\begin{tabular}{lc}
\hline\hline
Material & $D$ (eV)\\
\hline
$h$-BN       & 0.96~\cite{Kores-correct2017} \\
Silicene   & 0.7~\cite{Croy2020} \\
Graphene   & 1.36--1.4~\cite{Zhang2015,Croy2020} \\
MoS$_2$    & 9.0~\cite{Croy2020} \\
InSe       & 23.1~\cite{Croy2020} \\
Diamane    & 47--48~\cite{Wu2021,Zheng2023} \\
B$_2$O$_2$ & 30.2 (this work) \\
\hline\hline
\end{tabular}
\end{table} 

\subsection{Mechanical and finite-temperature stability}
\label{sec:mechanical}

The in-plane elastic moduli were extracted from linear fits to the calculated stress-strain curves over the small strain interval $0\leq\varepsilon\leq0.02$ (Fig. S3(a)). 
The uniaxial moduli along the $x$ and $y$ directions are defined as

\begin{equation}
E_x=
\left. 
\frac{d\sigma_x}{d\varepsilon_x}
\right|_{\varepsilon_x\rightarrow0};
\qquad
E_y=
\left. \frac{d\sigma_y}{d\varepsilon_y}\right|_{\varepsilon_y\rightarrow0} 
\label{eq:uniaxial_moduli}
\end{equation}

For equibiaxial deformation, the same strain is applied simultaneously along both in-plane directions, $\varepsilon_x=\varepsilon_y=\varepsilon_{\mathrm{biaxial}}$, and the biaxial modulus is defined as
\begin{equation}
E_{\mathrm{biaxial}}
=
\left. 
\frac{d\sigma_{\mathrm{biaxial}}}
     {d\varepsilon_{\mathrm{biaxial}}}
\right|_{\varepsilon_{\mathrm{biaxial}}\rightarrow0};
\qquad
\sigma_{\mathrm{biaxial}}
=
\frac{\sigma_x+\sigma_y}{2} 
\label{eq:biaxial_modulus}
\end{equation}
Here, $\sigma_x$ and $\sigma_y$ are the in-plane normal stresses along the $x$ and $y$ directions, respectively, while $\sigma_{\rm biaxial}$ is their average under equibiaxial deformation. Similarly, $\epsilon_x$ and $\epsilon_y$ denote the corresponding uniaxial strains, whereas $\epsilon_{\rm biaxial}$ is the common strain applied along both in-plane directions. Linear fitting gives $E_x=557.84$ GPa, $E_y=564.41$ GPa, and $E_{\rm biaxial}=680.01$ GPa.\\
 \indent No substantial reconstruction of the B$_2$O$_2$ framework was
observed during the 38 ps \textit{ab initio} molecular dynamics (AIMD) simulations~\cite{NS1984} at 300 and 1000 K. The simulations employed a $4\times4\times1$ supercell, and a time step of 1 fs. As shown in Fig. S3(b), the total energies exhibit bounded fluctuations at both temperatures. The initial and final configurations of the 1000 K simulation are compared in Figs. S3(c,d), respectively. The B$_2$O$_2$ framework remains intact without substantial reconstruction, supporting its kinetic stability at elevated temperature.
\renewcommand{\thefigure}{S\arabic{figure}} 
\begin{figure*}[t]  
\centering
\includegraphics[width=0.7\textwidth]{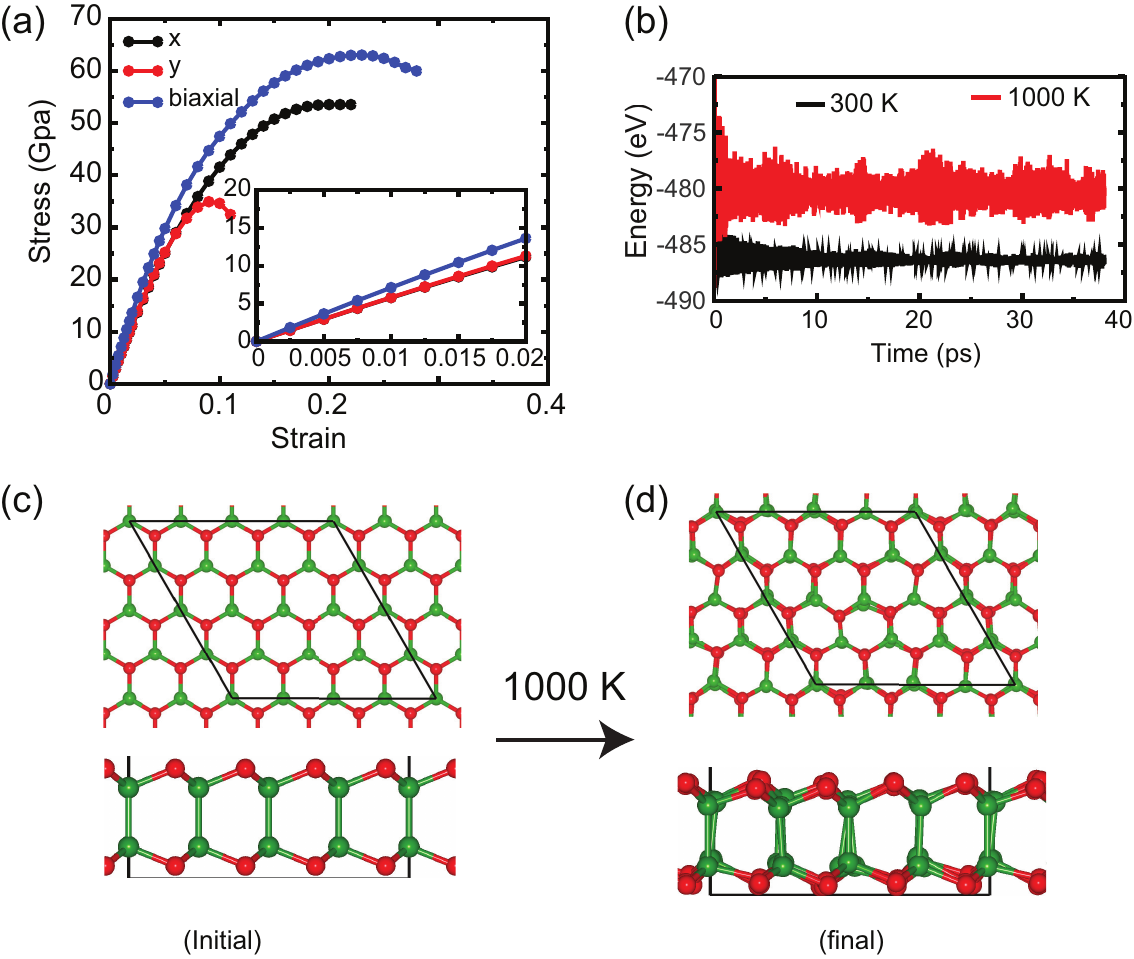}
\caption {(a) Stress-strain responses under uniaxial deformation along the $x$ and $y$ directions and under equibiaxial deformation. The inset shows the small strain interval, $0\leq\varepsilon\leq0.02$, used for linear fitting and extraction of $E_x$, $E_y$, and $E_{\mathrm{biaxial}}$. (b) Evolution of the total energy during \textit{ab initio} molecular dynamics simulations at 300 and 1000 K over 38 ps. (c,d) Top and side views of the initial and final B$_2$O$_2$ configurations, respectively, for the AIMD simulation at 1000 K. The B$_2$O$_2$ framework remains intact without noticeable reconstruction, supporting its kinetic stability at elevated temperature.}
\end{figure*} 

\subsection{Electronic properties}
\label{sec:electronic}

The electron localization function (ELF) shown in Fig. S4(a,b) reveals pronounced charge localization around the O atoms together with directional B--O bonding within the corrugated B$_2$O$_2$ framework. The corresponding electronic band structure and atom and orbital projected density of states (Fig. S4(c)) indicate an indirect band gap of $4.74$ eV, with the valence band maximum located along the $\Gamma$-K path and the conduction band minimum at M.\\
The valence band edge is dominated primarily by O-$p_{z}$ states, whereas the conduction band edge is mainly composed of B-$p_{z}$ states, indicating distinct orbital characters on the two sides of the band gap. 
\renewcommand{\thefigure}{S\arabic{figure}} 
\begin{figure*}[!ht] 
\centering
\includegraphics[width=0.7\textwidth]{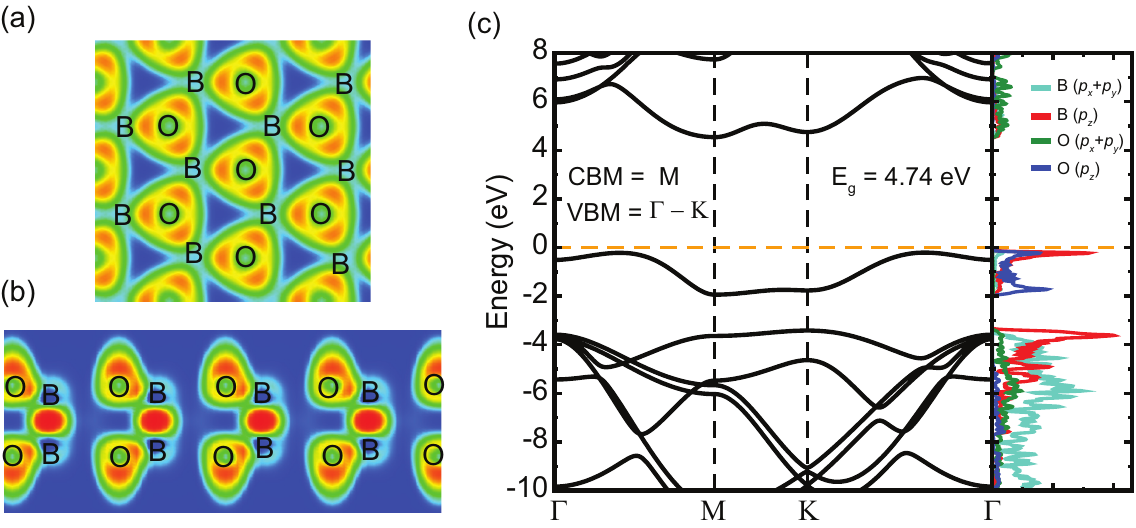}
\caption{(a-b) Top and side views of the electron localization function (ELF) of the corrugated $\mathrm{B_2O_2}$ monolayer, illustrating electron localization around the O atoms and directional B-O bonding. (c) Electronic band structure together with the atom and orbital-projected density of states (PDOS).}  
\end{figure*}
\\
\\
\section{Phonon symmetry, polarization, and mirror parity}
\label{sec:symmetry}
Harmonic phonon frequencies and eigenvectors were obtained using the finite displacement method implemented in Phonopy~\cite{togo2015}, with a $3\times3\times1$ electronic $\mathbf{k}$-point mesh for the displaced structures. The converged IFC2 from the $12\times12\times1$ supercell (Sec. S1.2 and Fig. S2) was used consistently for the phonon dispersion, polarization, mirror parity, and thermal transport analyses.

\subsection{$\Gamma$-point symmetry and polarization}
\label{sec:gamma_symmetry}

To verify the symmetry and polarization characters of the phonon modes discussed in the main text, we performed an irreducible representation analysis at $\Gamma$, whose little group is $D_{3h}$. The dynamical matrix constructed from the second order interatomic force constants was diagonalized according to
\begin{equation}
D(\mathbf q)\mathbf e_{\nu}(\mathbf q)
= \omega_{\nu}^{2}(\mathbf q)\mathbf e_{\nu}(\mathbf q) 
\end{equation}
where $\omega_{\nu}(\mathbf q)$ and $\mathbf e_{\nu}(\mathbf q)$ are the phonon frequency and normalized eigenvector, respectively, of branch $\nu$ at wave vector $\mathbf q$. At $\Gamma$ ($\mathbf q=\mathbf 0$), $e_{j\alpha}^{(\nu)}$ denotes the displacement amplitude of atom $j$ along Cartesian direction $\alpha$ for branch $\nu$. The normalized in-plane and out-of-plane eigenvector projection weights were defined as

\begin{equation}
P_{xy}^{(\nu)}
=
\sum_j
\left[
\left|e_{jx}^{(\nu)}\right|^2+
\left|e_{jy}^{(\nu)}\right|^2
\right],
\qquad
P_z^{(\nu)}
=
\sum_j
\left|e_{jz}^{(\nu)}\right|^2 
\label{eq:phonon_projection}
\end{equation}
with $P_{xy}^{(\nu)}+P_z^{(\nu)}=1$. Modes with $P_{xy}^{(\nu)}\simeq1$ and $P_z^{(\nu)}\simeq1$ are classified as predominantly in-plane and out-of-plane polarized, respectively. The $\mathbf q$-dependent out-of-plane projection $P_z^{(\nu)}(\mathbf q)$ was used to color code the polarization resolved phonon dispersion, as shown in Fig. 1(d) of the main text. Table S3 reports both $P_{xy}^{(\nu)}$ and $P_z^{(\nu)}$ at $\Gamma$, together with the corresponding frequencies, irreducible representations, horizontal mirror parities, and mode characters. 
\begin{table*}[!ht] 
\centering
\caption{Frequencies, irreducible representations, horizontal mirror parities, normalized Cartesian eigenvector projection weights $P_{xy}^{(\nu)}$ and $P_z^{(\nu)}$, and corresponding mode characters of the phonons at $\Gamma$. Horizontal mirror parity is assigned from the character of $\sigma_h$ in the corresponding $D_{3h}$ irreducible representation: primed representations are even (+1), whereas double primed representations are odd (-1). The predominantly in-plane or out-of-plane polarization character is determined independently from $P_{xy}^{(\nu)}$ and $P_z^{(\nu)}$, respectively.}  
\begin{ruledtabular}
\begin{tabular}{ccccccc}
Mode index at $\Gamma$ &
Frequency (THz) &
Irrep &
$\sigma_h$ parity &
$P_{xy}^{(\nu)}$ & 
$P_z^{(\nu)}$ & 
Mode character
\\
\hline
1,2  & 0.000    & $E'$     & even & $0.9989,\ 0.9998$ & $0.0011,\ 0.0002$ & TA/LA acoustic doublet \\
3    & 0.000001 & $A_2''$  & odd  & 0.0013 & 0.9987 & flexural acoustic (ZA) \\
4,5  & 9.290    & $E''$    & odd  & 1.0000 & 0.0000 & in-plane optical doublet \\
6    & 14.052   & $A_1'$   & even & 0.0000 & 1.0000 & out-of-plane optical \\
7,8  & 24.884   & $E'$     & even & 1.0000 & 0.0000 & in-plane optical doublet \\
9,10 & 25.689   & $E''$    & odd  & 1.0000 & 0.0000 & in-plane optical doublet \\
11   & 29.822   & $A_2''$  & odd  & 0.0000 & 1.0000 & out-of-plane optical \\
12   & 34.474   & $A_1'$   & even & 0.0000 & 1.0000 & out-of-plane optical \\
\end{tabular}
\end{ruledtabular}
\end{table*}
\\
Collecting all 12 phonon branches listed in Table S3, the complete $\Gamma$-point vibrational representation is $\Gamma_{\mathrm{vib}}  =2A_1'\oplus2A_2''\oplus2E'\oplus2E''$. Separating the acoustic translations gives $\Gamma_{\mathrm{ac}}=A_2''\oplus E'$, comprising the odd $A_2''$ flexural mode (ZA) and the even $E'$ in-plane translational doublet (TA/LA). The remaining optical representation is $\Gamma_{\mathrm{opt}} =2A_1'\oplus A_2''\oplus E'\oplus2E''$.

\subsection{Finite-$q$ mirror parity}
\label{sec:finiteq_parity}

Mirror parity and Cartesian polarization are distinct classifications. For example, the $E''$ optical modes in Table S3 are mirror odd but in-plane polarized, whereas the $A_1'$ modes are mirror even but out of plane polarized. This distinction arises because horizontal reflection transforms both the displacement components and the symmetry related atomic sites. At finite wave vector, we evaluate the mirror expectation value of each normalized phonon eigenvector as
$\langle\sigma_h\rangle_{\mathbf q\nu}
=\mathbf e_{\mathbf q\nu}^{\dagger}
D_{\sigma_h}(\mathbf q)\mathbf e_{\mathbf q\nu}$
where $D_{\sigma_h}(\mathbf q)$ includes the Cartesian reflection, atomic permutation, and corresponding Bloch phases. Values of +1 and -1 identify mirror even and mirror odd modes, respectively. Table S4 confirms that ZA is mirror odd and TA/LA are mirror even at the five sampled long wavelength points along $\Gamma$-M. Frequency ordered indices do not define global branch identities because branches can cross and reorder. The scattering analysis therefore uses the explicitly evaluated mirror eigenvalue of every participating acoustic and optical mode. Continuous branch tracking and the extended ZA scattering analysis along $\Gamma$-M are described in Sec. S4.1.

\begin{table}[H]
\caption{Calculated horizontal mirror eigenvalues of all twelve phonon modes at five long wavelength points along $\Gamma$-M, with $\mathbf q=(q_x,0,0)$ in reciprocal lattice coordinates. Modes are ordered independently by increasing frequency at each wave vector; ZA is the lowest frequency mode at these points. The interchange of parity entries at indices 8 and 10 is consistent with frequency reordering near a branch crossing and does not indicate parity mixing.}   
\begin{ruledtabular}
\begin{tabular}{c@{\hspace{1.2cm}}ccccc}
&
\multicolumn{5}{c}{
Mirror parity along $\Gamma$-$M$,
$\mathbf q=(q_x,0,0)$
} 
\\
\cline{2-6}
Frequency ordered index
&
$q_x=0.01875$
&
$q_x=0.02500$
&
$q_x=0.03125$
&
$q_x=0.03750$
&
$q_x=0.04375$
\\
\hline
1 (ZA) & $-1$ & $-1$ & $-1$ & $-1$ & $-1$ \\
2      & $+1$ & $+1$ & $+1$ & $+1$ & $+1$ \\
3      & $+1$ & $+1$ & $+1$ & $+1$ & $+1$ \\
4      & $-1$ & $-1$ & $-1$ & $-1$ & $-1$ \\
5      & $-1$ & $-1$ & $-1$ & $-1$ & $-1$ \\
6      & $+1$ & $+1$ & $+1$ & $+1$ & $+1$ \\
7      & $+1$ & $+1$ & $+1$ & $+1$ & $+1$ \\
8      & $+1$ & $-1$ & $-1$ & $-1$ & $-1$ \\
9      & $-1$ & $-1$ & $-1$ & $-1$ & $-1$ \\
10     & $-1$ & $+1$ & $+1$ & $+1$ & $+1$ \\
11     & $-1$ & $-1$ & $-1$ & $-1$ & $-1$ \\
12     & $+1$ & $+1$ & $+1$ & $+1$ & $+1$ \\ 
\end{tabular}
\end{ruledtabular}
\end{table} 

\section{Lattice thermal-transport}
\label{sec:thermal}
\subsection{Anharmonic force constants and BTE calculations}
\label{sec:bte}

The harmonic IFC2 obtained using the converged $12\times12\times1$ supercell was adopted following comparisons with $8\times8\times1$ and $10\times10\times1$ supercells (Sec. S2 and Fig. S2). The anharmonic third order interatomic force constants (IFC3) were generated using the finite displacement method with a $6\times6\times1$ supercell. Interaction cutoffs of 5 and 6~\AA{} were tested, and 6~\AA{} was adopted as sufficient. The phonon Boltzmann transport equation was solved iteratively using ShengBTE~\cite{Sheng2014}, including anharmonic three phonon scattering and naturally occurring isotope disorder. Convergence with respect to Brillouin zone sampling was examined using phonon meshes ranging from $80\times80\times1$ to $180\times180\times1$. The lattice thermal conductivity was nearly converged at $160\times160\times1$, which was adopted for the final calculations. The resulting lattice thermal conductivity tensor is

\begin{equation}
\kappa^{\alpha\beta}
=
\frac{1}{k_{\mathrm B}T^2NV}
\sum_{\lambda}
(\hbar\omega_\lambda)^2
f_\lambda^0(1+f_\lambda^0)
v_\lambda^\alpha F_\lambda^\beta 
\end{equation}

where $\lambda\equiv(\mathbf{q},\nu)$ labels a phonon mode with wave vector $\mathbf{q}$ and branch index $\nu$, and $\alpha$ and $\beta$ denote Cartesian directions. Here, $k_{\mathrm B}$ is the Boltzmann constant, $\hbar$ is the reduced Planck constant, $V$ is the unit cell volume, and $N$ is the number of sampled wave vectors. The quantities $\omega_\lambda$, $f_\lambda^0$, and $\mathbf{v}_\lambda$ are the angular frequency, equilibrium Bose-Einstein occupation, and group velocity of mode $\lambda$, respectively. The generalized mean free displacement function $F_\lambda^\beta$ is obtained by solving the linearized BTE iteratively~\cite{Bro2007,Lindsay2008,Ward2009,Ward2010} and is written as $F_\lambda^\beta=\tau_\lambda^0 \left(v_\lambda^\beta+\Delta_\lambda^\beta\right)$, where $\tau_\lambda^0$ is the single mode relaxation time and $\Delta_\lambda^\beta$ is the iterative correction arising from coupling to the nonequilibrium populations of other phonon modes.

For three phonon processes, the triplet resolved absorption $(\lambda+\lambda'\leftrightarrow\lambda'')$ and emission $(\lambda\leftrightarrow\lambda'+\lambda'')$ rates are expressed as~\cite{Lindsay2009}
 
\begin{equation}
P_{\lambda\lambda'\lambda''}^{+}
=
\frac{\pi\hbar}{4}
\left(f_{\lambda'}^0-f_{\lambda''}^0\right)
\frac{
\delta(\omega_\lambda+\omega_{\lambda'}-\omega_{\lambda''})
}{
\omega_\lambda\omega_{\lambda'}\omega_{\lambda''}
}
\left|\Phi_{\lambda\lambda'\lambda''}^{+}\right|^2 
\end{equation}
\begin{equation}
P_{\lambda\lambda'\lambda''}^{-}
=
\frac{\pi\hbar}{4}
\left(f_{\lambda'}^0+f_{\lambda''}^0+1\right)
\frac{
\delta(\omega_\lambda-\omega_{\lambda'}-\omega_{\lambda''})
}{
\omega_\lambda\omega_{\lambda'}\omega_{\lambda''}
}
\left|\Phi_{\lambda\lambda'\lambda''}^{-}\right|^2 
\end{equation}

Here, $\Phi^{\pm}_{\lambda\lambda'\lambda''}$ are the mass weighted three phonon interaction matrix elements for absorption and emission, respectively. They describe the anharmonic coupling among the three participating modes and are obtained by contracting the third order IFCs with their phonon eigenvectors, including the appropriate Bloch phases and process dependent complex conjugation. The standard expressions implemented in ShengBTE~\cite{Sheng2014} were used. Isotope scattering rates were evaluated using the standard mass disorder expression with natural isotopic abundances, as implemented in the same code. 
\\
\renewcommand{\thefigure}{S\arabic{figure}} 
\begin{figure*}[!ht]  
\centering 
\includegraphics[width=0.8\textwidth]{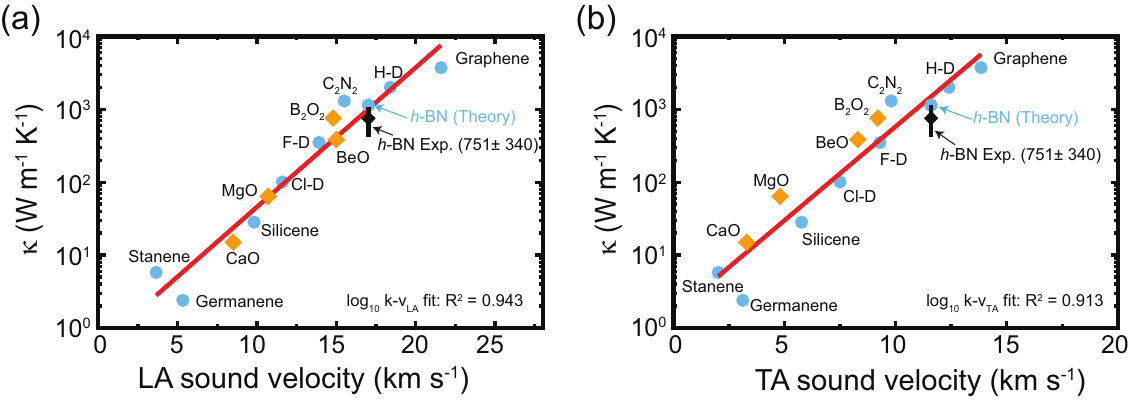}
\caption{Lattice thermal conductivity ($\kappa$) and acoustic sound velocities of selected two-dimensional materials. Lattice thermal conductivity $\kappa$ versus (a) longitudinal acoustic (LA) and (b) transverse acoustic (TA) sound velocity. Orange diamonds denote B$_2$O$_2$ and reference oxides~\cite{Mor2021}; blue sphere denote elemental monolayers~\cite{Pen2016}, $h$-BN~\cite{Fan2019}, diamanes~\cite{Mos2020}, and C$_{2}$N$_{2}$~\cite{Chen2025}. D denotes diamane. The black diamond and vertical error bar in each panel show the experimental $h$-BN conductivity, 
$751\pm340$ $\mathrm{W\,m^{-1}\,K^{-1}}$~\cite{Qi2019}. Red lines are linear fits of  $\log_{10}\kappa$ versus the corresponding sound velocity, with $R^2=0.943$ and $0.913$ for LA and TA, respectively.} 
\label{fig:kappa_velocity}    
\end{figure*} 
\\
\\
\subsection{Correlation between acoustic velocities and thermal conductivity}
Across the selected two-dimensional materials, the lattice thermal conductivity ($\kappa$) correlates positively with the LA and TA sound velocities, with linear fits of $\log_{10}\kappa$ yielding $R^2=0.943$ and $0.913$, respectively (Fig. S5(a,b). Compared with BeO, B$_2$O$_2$ exhibits a similar LA sound velocity but a higher TA sound velocity and approximately twice the thermal conductivity. B$_2$O$_2$ also lies above both fitted trends, indicating that its high thermal conductivity is not fully captured by the in-plane acoustic sound velocities alone. Phonon lifetimes and branch resolved contributions must therefore also be considered.
\\
\\
\\
\\
\\
\\
\\
\\
\subsection{Thickness normalization} 
\label{sec:thickness}
The thermal conductivity reported by ShengBTE is defined using the three dimensional simulation-cell volume and therefore depends on the out-of-plane cell length $L_z$. We express the monolayer thermal conductivity using an effective thickness $t_{\mathrm{eff}}$:

\begin{equation}
\kappa = \kappa_{\mathrm{cell}} \frac{L_z}{t_{\mathrm{eff}}} 
\label{eq:thickness_normalization}
\end{equation}
where $\kappa_{\mathrm{cell}}$ is the thermal conductivity calculated using the simulation cell volume. To obtain a physically motivated effective thickness, we constructed and optimized a layered bulk B$_2$O$_2$ reference structure in the present work. The calculated interlayer gap is approximately 2.96~\AA{}. Combining this gap with the geometric monolayer thickness of 2.924~\AA{} gives a layer repeat distance of approximately 5.88~\AA{}, which we adopt as $t_{\mathrm{eff}}$. This effective thickness includes both the corrugated monolayer and the gap between adjacent layers. At T = 300 K, using $\kappa_{\mathrm{cell}} = 222.32 \mathrm{W\,m^{-1}\,K^{-1}}$, $L_z=20$~\AA{}, and $t_{\mathrm{eff}}=5.88$~\AA{}, we obtain 
$\kappa = 222.32\times\frac{20}{5.88}$ $\simeq$ 756 $\mathrm{W\,m^{-1}\,K^{-1}}$. 

\renewcommand{\thefigure}{S\arabic{figure}} 
\begin{figure*}[!ht]  
\centering
\includegraphics[width=0.8\textwidth]{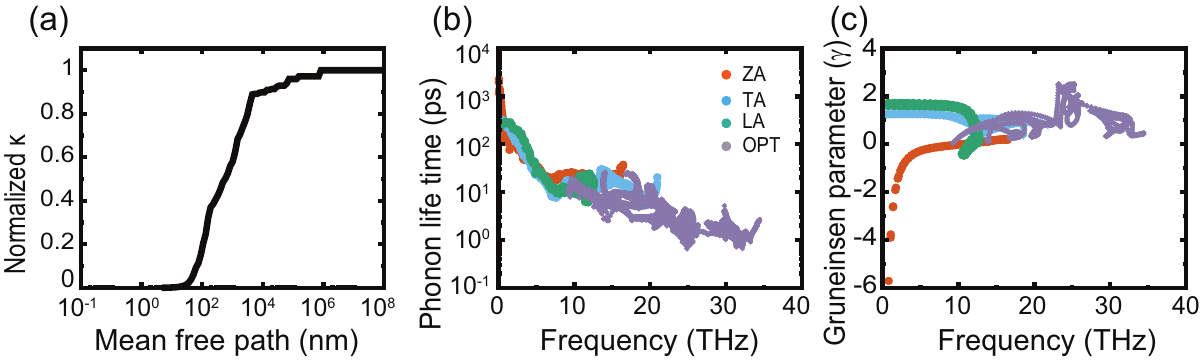}
\caption {(a) Normalized cumulative lattice thermal conductivity as a function of phonon mean free path at 300 K. (b) Phonon lifetime versus phonon frequency for the longitudinal acoustic (LA), transverse acoustic (TA), flexural acoustic (ZA), and optical (OPT) branches. (c) Mode Gr\"uneisen parameters ($\gamma$), as a function of phonon frequency. ZA, TA, LA, and OPT denote the flexural acoustic, in-plane transverse acoustic, longitudinal acoustic, and optical branches, respectively.}  
\end{figure*} 

\section{Acoustic scattering-channel analysis}
\subsection{Acoustic--optical decomposition and mirror-parity-resolved ZA scattering} 
To resolve the three phonon scattering of the ZA branch without relying on local frequency ordering, the three acoustic branches were continuously tracked over the $160\times160\times1$ phonon mesh using maximum eigenvector overlap matching. The resulting branch identities are denoted by $Z$, $T$, and $L$ for ZA, TA, and LA, respectively, while $O$ collectively denotes the remaining nine optical modes at each wave vector. The label $O$ is used only for channel bookkeeping; optical modes are treated individually when their mirror eigenvalues are evaluated. We first consider a representative long wavelength ZA mode at $\mathbf q=(0.01875,0,0)$ and $T=300$ K. Wave vectors are expressed in reciprocal lattice coordinates.

For a fixed initial mode $\lambda$ on the tracked ZA branch $Z$, the two partner modes are classified by $\beta,\gamma\in\{Z,T,L,O\}$, giving 16 ordered channels $Z\beta\gamma$. The contribution of each channel to the three phonon scattering rate is
\begin{equation}
\Gamma_{\lambda}^{Z\beta\gamma}=\sum_{\lambda',\lambda''}w_{\lambda\lambda'\lambda''}\Gamma_{\lambda\lambda'\lambda''}\delta_{\beta}(\lambda')\delta_{\gamma}(\lambda'')
\end{equation}
where $w_{\lambda\lambda'\lambda''}$ is the irreducible triplet multiplicity, 

$\Gamma_{\lambda\lambda'\lambda''}$ is the triplet resolved contribution to the inverse three phonon lifetime, including absorption and emission processes in the phono3py convention. The indicator functions $\delta_{\beta}$ and $\delta_{\gamma}$ select the corresponding tracked partner mode classes. The sum is taken over irreducible momentum conserving triplets and their partner bands, with $w_{\lambda\lambda'\lambda''}$ restoring the triplet multiplicities in the underlying phono3py convention. Summing all partner classes gives
\begin{equation}
\Gamma_{\lambda}^{\mathrm{tot}}=\sum_{\beta,\gamma}\Gamma_{\lambda}^{Z\beta\gamma}=2.6397354\times10^{-4}\ \mathrm{ps}^{-1} 
\end{equation}
Here $\Gamma$ denotes the inverse lifetime in $\mathrm{ps}^{-1}$, obtained from the phono3py half-linewidth $g$ in ordinary frequency THz using $\Gamma\,[\mathrm{ps}^{-1}]=4\pi g\,[\mathrm{THz}]$. Thus $\Gamma_{\lambda}^{\mathrm{tot}}=(\tau_{\lambda}^{3\mathrm{ph}})^{-1}$. This convention applies to every triplet and channel contribution as well as their total. 

The complete channel decomposition is given in Table~\ref{tab:S5}. Grouping the two partner modes by their acoustic ($A\in\{Z,T,L\}$) or optical ($O$) character yields
\begin{equation}
ZAA=79.2497\%,\qquad ZAO=0.8499\%,\qquad ZOO=19.9004\% 
\end{equation}
where $ZAA$, $ZAO$, and $ZOO$ correspond to the AA, AO, and OO sectors, respectively. AA denotes two acoustic partners, AO one acoustic and one optical partner, and OO two optical partners; $ZAO$ includes both acoustic-optical partner orderings. For this long wavelength ZA mode, channels with two acoustic partners dominate, while channels with two optical partners provide a substantial secondary contribution. Because the optical manifold contains both mirror even and mirror odd modes, the coarse $ZAO$ and $ZOO$ classes cannot be assigned a unique mirror parity. Each triplet is classified by the product of its horizontal mirror eigenvalues,
\begin{equation}
m_{\lambda}m_{\lambda'}m_{\lambda''}=\pm1 
\end{equation}
where $m_{\lambda}=\pm1$ denotes the mirror eigenvalue of the corresponding phonon mode. The horizontal mirror selection rule permits triplets with product $+1$ and forbids those with product $-1$. Their scattering rate contributions are accumulated separately as $\Gamma_+$ and $\Gamma_-$, respectively. For the representative mode, the parity-resolved scattering rates are
\begin{equation}
\Gamma_+=2.639735443\times10^{-4}\ \mathrm{ps}^{-1},\qquad \Gamma_-=6.876475120\times10^{-26}\ \mathrm{ps}^{-1}
\end{equation}
corresponding to
\begin{equation}
\frac{\Gamma_-}{\Gamma_{\lambda}^{\mathrm{tot}}}\times100=2.60\times10^{-20}\% 
\end{equation}
The mirror forbidden contribution is thus negligible within numerical precision. In the acoustic-acoustic sector, the total scattering rate is $\Gamma_{AA}=2.091981833\times10^{-4}\ \mathrm{ps}^{-1}$, dominated by the four mirror allowed channels $ZZT$, $ZZL$, $ZTZ$, and $ZLZ$. The five mirror forbidden acoustic channels $ZZZ$, $ZTT$, $ZTL$, $ZLT$, and $ZLL$ sum to approximately $6.80\times10^{-26}\ \mathrm{ps}^{-1}$.

To test whether this suppression is restricted to the representative long wavelength mode, the same exact triplet parity analysis was performed for tracked ZA modes along $\Gamma$-M, from $q_x=0.01875$ to the zone boundary M at $q_x=0.50$. The ZA identity was followed continuously by eigenvector overlap, even where its local frequency order index changes at larger wave vectors. The results are summarized in Table S6. Over all sampled points, the forbidden fraction remains at the level of numerical residuals:
\begin{equation}
\max_{\mathbf q}\left[\frac{\Gamma_-(\mathbf q)}{\Gamma^{\mathrm{tot}}(\mathbf q)}\times100\right]=1.03\times10^{-19}\%
\end{equation}
The horizontal mirror selection rule is therefore satisfied to numerical precision throughout the sampled $\Gamma$-M ZA branch. The acoustic-optical composition of the allowed scattering nevertheless varies considerably along the branch. At $q_x=0.01875$, the coarse contributions are
\begin{equation}
ZAA=79.25\%,\qquad ZAO=0.85\%,\qquad ZOO=19.90\% 
\end{equation}
whereas at $q_x=0.40$ they become
\begin{equation}
ZAA=9.53\%,\qquad ZAO=56.90\%,\qquad ZOO=33.57\% 
\end{equation}
Thus, the relative weights of acoustic-acoustic, acoustic-optical, and optical-optical channels evolve substantially along the ZA branch. This redistribution is compatible with changes in the available three phonon phase space, interaction matrix elements, and mode character; the present decomposition alone does not isolate their individual effects. The mirror forbidden scattering rate remains negligible across the sampled branch, irrespective of the changing acoustic-optical composition of the allowed channels, consistent with the horizontal mirror selection rule. 

\begin{table}[H]
\caption{Three phonon scattering rate decomposition of the tracked ZA mode at $\mathbf{q}=(0.01875,0,0)$ and $T=300$ K. The labels $Z$, $T$, and $L$ denote the tracked ZA, TA, and LA branches, respectively, while $O$ denotes the remaining nine optical modes. The AA, AO, and OO sectors correspond to two acoustic partners, one acoustic and one optical partner, and two optical partners, respectively. All percentages are normalized to $\Gamma_{\lambda}^{\mathrm{tot}} =2.6397354\times10^{-4} \mathrm{ps}^{-1}$. The last two columns give the mirror allowed ($+$) and mirror forbidden ($-$) scattering rate contributions. Tiny nonzero forbidden contributions are numerical residuals.}
\label{tab:S5}
\centering 
\begin{ruledtabular}
\begin{tabular}{cccccc}
\hline
Channel & Sector
& $\Gamma_{\lambda}^{Z\beta\gamma}$ ($\mathrm{ps}^{-1}$)
& Contribution (\%)
& $\Gamma_{+}$ ($\mathrm{ps}^{-1}$)
& $\Gamma_{-}$ ($\mathrm{ps}^{-1}$)\\
\hline
ZZZ & AA & $6.795\times10^{-26}$ & 0.000000
    & $0$ & $6.795\times10^{-26}$ \\
ZZT & AA & $9.533\times10^{-5}$ & 36.112850
    & $9.533\times10^{-5}$ & $0$ \\
ZZL & AA & $7.874\times10^{-6}$ & 2.982798
    & $7.874\times10^{-6}$ & $0$ \\
ZZO & AO & $0$ & 0.000000
    & $0$ & $0$ \\
ZTZ & AA & $7.201\times10^{-5}$ & 27.277533
    & $7.201\times10^{-5}$ & $0$ \\
ZTT & AA & $5.582\times10^{-30}$ & 0.000000
    & $0$ & $5.582\times10^{-30}$ \\
ZTL & AA & $5.137\times10^{-31}$ & 0.000000
    & $0$ & $5.137\times10^{-31}$ \\
ZTO & AO & $7.721\times10^{-7}$ & 0.292477
    & $7.721\times10^{-7}$ & $0$ \\
ZLZ & AA & $3.399\times10^{-5}$ & 12.876497
    & $3.399\times10^{-5}$ & $0$ \\
ZLT & AA & $3.805\times10^{-30}$ & 0.000000
    & $0$ & $3.805\times10^{-30}$ \\
ZLL & AA & $2.720\times10^{-29}$ & 0.000000
    & $0$ & $2.720\times10^{-29}$ \\
ZLO & AO & $0$ & 0.000000
    & $0$ & $0$ \\
ZOZ & AO & $0$ & 0.000000
    & $0$ & $0$ \\
ZOT & AO & $1.949\times10^{-7}$ & 0.073834
    & $1.949\times10^{-7}$ & $0$ \\
ZOL & AO & $1.277\times10^{-6}$ & 0.483631
    & $1.277\times10^{-6}$ & $0$ \\
ZOO & OO & $5.253\times10^{-5}$ & 19.900380
    & $5.253\times10^{-5}$ & $7.737\times10^{-28}$ \\
\end{tabular}
\end{ruledtabular} 
\end{table}

\begin{table}[H]
\caption{Wave vector dependence of the mirror parity resolved three phonon scattering rate of the tracked ZA branch along $\Gamma$-M at T = 300 K. The wave vectors $\mathbf{q}=(q_x,0,0)$ are given in reciprocal lattice coordinates, with M represented by $q_x=0.50$. $\nu_{\mathrm{ZA}}$ is the ordinary phonon frequency in THz. The last column gives the percentage of the total ZA scattering rate arising from triplets with $m_{\lambda}m_{\lambda'}m_{\lambda''}=-1$.}
\label{tab:S6}
\centering
\begin{ruledtabular}
\begin{tabular}{cccc}
$q_x$
& $\nu_{\mathrm{ZA}}$ (THz)
& $\Gamma^{\mathrm{tot}}$ ($\mathrm{ps}^{-1}$)
& $\Gamma_{-}/\Gamma^{\mathrm{tot}}$ (\%) \\
\hline
0.01875 &  0.0630 & $2.640\times10^{-4}$ & $2.60\times10^{-20}$ \\
0.02500 &  0.1364 & $5.608\times10^{-4}$ & $1.03\times10^{-19}$ \\
0.05000 &  0.5922 & $3.296\times10^{-3}$ & $1.34\times10^{-21}$ \\
0.07500 &  1.2540 & $8.205\times10^{-3}$ & $1.09\times10^{-22}$ \\
0.10000 &  2.0444 & $7.578\times10^{-3}$ & $9.90\times10^{-24}$ \\
0.12500 &  2.9085 & $1.367\times10^{-2}$ & $4.34\times10^{-23}$ \\
0.15000 &  3.8122 & $2.130\times10^{-2}$ & $5.19\times10^{-23}$ \\
0.17500 &  4.7350 & $2.766\times10^{-2}$ & $8.71\times10^{-24}$ \\
0.20000 &  5.6644 & $4.135\times10^{-2}$ & $8.69\times10^{-22}$ \\
0.25000 &  7.5094 & $5.519\times10^{-2}$ & $5.97\times10^{-23}$ \\
0.30000 &  9.2927 & $4.052\times10^{-2}$ & $5.81\times10^{-24}$ \\
0.35000 & 10.9556 & $4.732\times10^{-2}$ & $5.12\times10^{-23}$ \\
0.40000 & 12.4086 & $7.184\times10^{-2}$ & $2.19\times10^{-22}$ \\
0.45000 & 13.4853 & $6.102\times10^{-2}$ & $1.35\times10^{-21}$ \\
0.50000 & 13.9060 & $4.995\times10^{-2}$ & $7.03\times10^{-23}$ \\
\end{tabular}
\end{ruledtabular}
\end{table}

\subsection{Normal and Umklapp contributions}
\label{sec:normal_umklapp}

To complement the mirror parity resolved channel analysis, we further
decompose the acoustic mode three phonon scattering rates into Normal (N) and Umklapp (U) contributions at $T=300$ K. Here, $\Gamma_{\mathrm{tot}}=\Gamma_N+\Gamma_U$, where $\Gamma_N$ and $\Gamma_U$ denote the Normal and Umklapp contributions, respectively. Normal processes conserve crystal momentum without the involvement of a nonzero reciprocal lattice vector, whereas Umklapp processes involve a nonzero reciprocal lattice vector and therefore provide a momentum relaxing scattering channel. The long wavelength ZA, TA, and LA modes are overwhelmingly dominated by N processes throughout the sampled near $\Gamma$ region along $\Gamma$-M, with the U contribution remaining below approximately $3\%$. 

\begin{table}[H]
\centering
\caption{Normal and Umklapp process contributions to the three phonon scattering rates of the long wavelength acoustic modes of monolayer B$_2$O$_2$ at $T=300$ K. The wave vectors are sampled along $\Gamma$-M as $\mathbf{q}=(q_x,0,0)$ in reciprocal lattice coordinates. The fractional contributions are defined as $N(\%)=100\,(\Gamma_N/\Gamma_{\mathrm{tot}})$ and $U(\%)=100\,(\Gamma_U/\Gamma_{\mathrm{tot}})$, with $\Gamma_{\mathrm{tot}}=\Gamma_N+\Gamma_U$.} 
\label{tab:S5_acoustic_channels}
\begin{ruledtabular}
\begin{tabular}{c cc cc cc}
$q_x$
& \multicolumn{2}{c}{ZA}
& \multicolumn{2}{c}{TA}
& \multicolumn{2}{c}{LA} \\
\cline{2-3}\cline{4-5}\cline{6-7}
& $N$ (\%) & $U$ (\%)
& $N$ (\%) & $U$ (\%)
& $N$ (\%) & $U$ (\%) \\
\hline
0.01875 & 99.954 & 0.046 & 99.823 & 0.177 & 99.786 & 0.214 \\
0.02500 & 99.752 & 0.248 & 99.403 & 0.597 & 99.426 & 0.574 \\
0.03125 & 98.214 & 1.786 & 99.039 & 0.961 & 98.057 & 1.943 \\
0.03750 & 99.114 & 0.886 & 97.827 & 2.173 & 97.060 & 2.940 \\
0.04375 & 97.287 & 2.713 & 97.009 & 2.991 & 97.008 & 2.992 \\
\end{tabular}
\end{ruledtabular}
\end{table}

\section{B$_2$O$_2$ on Cu(111) and Ni(111)}
\label{sec:substrates}

\subsection{Binding energetics and interfacial charge transfer}
\label{sec:interface}

To assess the feasibility of substrate assisted growth of monolayer B$_2$O$_2$, we investigated its interaction with Cu(111) and Ni(111) substrates using the DFT-D3 dispersion correction. Both Cu(111) and Ni(111) substrates were modeled using six layer slabs. The lattice mismatch between the B$_2$O$_2$ monolayer and the matched substrate surface cells was 2.7$\%$ for Cu(111) and 0.4$\%$ for Ni(111). Spin polarization was included in the Ni(111) interface calculations, whereas Cu(111) was found to be nonmagnetic. For each substrate, two adsorption configurations were considered, with either the B atoms (B-top) or the O atoms (O-top) of the B$_2$O$_2$ monolayer facing the metal surface. The binding energy of the $\mathrm{B_2O_2}$/metal interface was calculated as

\begin{equation}
E_{\mathrm{b}}
=
E_{\mathrm{B_2O_2/metal}}
-
E_{\mathrm{metal}}
-
E_{\mathrm{B_2O_2}} 
\end{equation}

where $E_{\mathrm{B_2O_2/metal}}$ is the total energy of the relaxed interface, $E_{\mathrm{metal}}$ is the total energy of the corresponding relaxed clean Cu(111) or Ni(111) slab, and $E_{\mathrm{B_2O_2}}$ is the total energy of the isolated $\mathrm{B_2O_2}$ monolayer constrained to the substrate matched in-plane lattice parameters and relaxed with respect to its internal atomic coordinates. A negative value of $E_{\mathrm{b}}$ indicates energetically favorable binding.

\renewcommand{\thetable}{S\arabic{table}} 
\begin{table}[h]
\caption{Calculated binding energies ($E_{\rm b}$), equilibrium interface separations ($d_{\rm inter}$), and net Bader charge transfer ($\Delta Q$) for the B$_2$O$_2$/Cu(111) and B$_2$O$_2$/Ni(111) interfaces. Positive $\Delta Q$ denotes electron transfer from the substrate to the B$_2$O$_2$ monolayer.}
\centering
\begin{tabular}{lccc}
\hline
Configuration & $E_{\rm b}$ (eV) & $d_{\rm inter}$ (\AA) & $\Delta Q$ ($e$)\\
\hline
Cu(111), B-top & $-0.15$ & 3.32 & +0.0130\\
Cu(111), O-top & $-0.16$ & 3.19 & +0.0132\\
Ni(111), B-top & $-0.125$ & 3.28 & +0.0096\\
Ni(111), O-top & $-0.14$ & 3.21 & +0.0116\\
\hline
\end{tabular}
\end{table}

\renewcommand{\thefigure}{S\arabic{figure}} 
\begin{figure*}[!ht] 
\centering 
\includegraphics[width=0.7\textwidth]{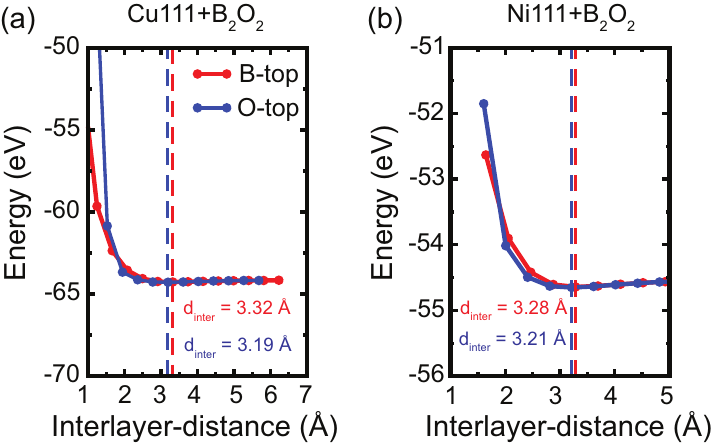} 
\caption {(a-b) Total energy as a function of the B$_2$O$_2$-substrate interlayer separation, $d_{\mathrm{inter}}$, for monolayer B$_2$O$_2$ on (a) Cu(111) and (b) Ni(111). Two interfacial registries are considered. In the B-top configuration, the interfacial B atoms of B$_2$O$_2$ face the surface Cu or Ni atoms, whereas in the O-top configuration, the interfacial O atoms face the surface metal atoms. The colored dashed vertical lines mark the equilibrium separations obtained from the minima of the corresponding total energy curves. The resulting equilibrium separations are 3.32 and 3.19~\AA{} for the B-top and O-top configurations on Cu(111), respectively, and 3.28 and 3.21~\AA{} for the corresponding configurations on Ni(111).} 
\end{figure*} 

The calculated binding energies, equilibrium interface separations, and Bader charge transfers are summarized in Table S8.  For both Cu(111) and Ni(111), the O-top configuration is energetically slightly more favorable than the corresponding B-top configuration, exhibiting slightly more negative binding energies and smaller equilibrium interface separations. Nevertheless, the differences between the two adsorption geometries are small. Bader charge analysis further reveals negligible charge transfer from the substrates to the B$_2$O$_2$ monolayer, amounting to approximately $0.0130e$ (B-top) and 0.0132$e$ (O-top) on Cu(111), and 0.0096$e$ (B-top) and 0.0116$e$ (O-top) on Ni(111). The consistently small Bader charge transfer indicates negligible net electron transfer across the interface, consistent with weak interfacial coupling. Taken together, the small binding energy magnitudes (0.125--0.16 eV), large equilibrium interface separations (3.19--3.32~\AA), and negligible interfacial charge transfer consistently demonstrate that B$_2$O$_2$ interacts only weakly with both Cu(111) and Ni(111).\\
 \indent The recent experimental realization of a two-dimensionalB$_2$O$_3$ monolayer on Pt(111) provides an important precedent for substrate assisted  growth of ordered boron oxide networks on metallic surfaces~\cite{T2025}. In that work, Pt(111) serves as the growth platform for the thermally driven assembly of crystalline B$_2$O$_3$, while the resulting monolayer interacts only weakly with the substrate, exhibiting a large interface separation (3.26~\AA), small adhesion energy, and negligible charge transfer~\cite{T2025}. Although our calculations indicate that the corresponding free standing B$_2$O$_3$ monolayer is dynamically unstable (structure 1; Fig. S1), its successful experimental realization demonstrates that a metal substrate can facilitate the formation of a two-dimensionalboron oxide phase even when the final overlayer is only weakly coupled to the support.

\renewcommand{\thefigure}{S\arabic{figure}} 
\begin{figure*}[!ht] 
\centering

\includegraphics[width=0.8\textwidth]{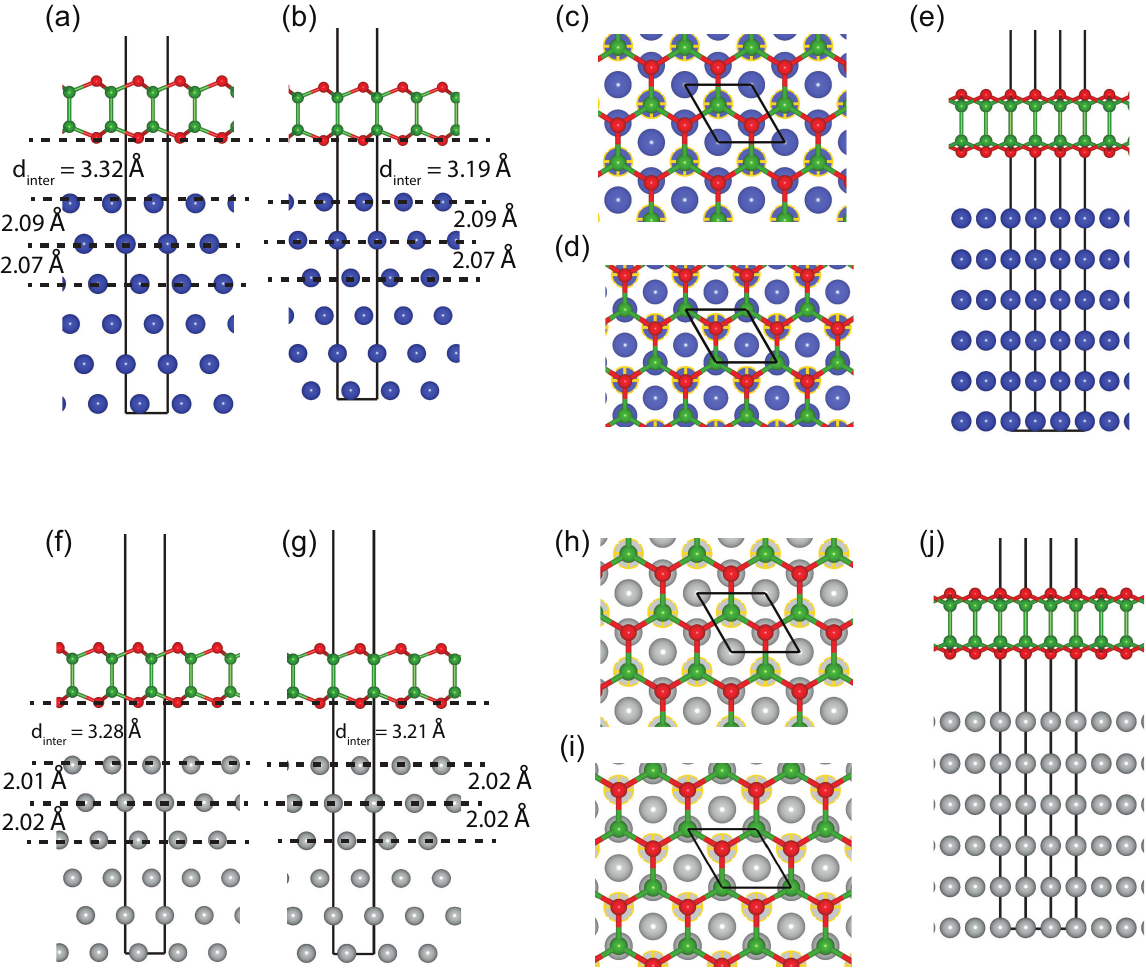} 
\caption {Optimized interfacial structures of monolayer B$_2$O$_2$ supported on Cu(111) and Ni(111). Panels (a-e) correspond to B$_2$O$_2$/Cu(111), while panels (f-j) correspond to B$_2$O$_2$/Ni(111). (a,c) Side and top views, respectively, of the B-top configuration on Cu(111), in which the interfacial B atoms of B$_2$O$_2$ face the surface Cu atoms. (b,d) Corresponding side and top views of the O-top configuration, in which the interfacial O atoms face the surface Cu atoms. The optimized interface separations are $d_{\mathrm{inter}}=3.32$ and 3.19~\AA{} for the B-top and O-top configurations, respectively. (e) Perspective view of the relaxed B$_2$O$_2$/Cu(111) interface. (f,h) Side and top views, respectively, of the B-top configuration on Ni(111), while (g,i) show the corresponding O-top configuration. The optimized interface separations are $d_{\mathrm{inter}}=3.28$ and 3.21~\AA{} for the B-top and O-top configurations on Ni(111), respectively. (j) Perspective view of the relaxed B$_2$O$_2$/Ni(111) interface. The surface metal atoms highlighted in yellow identify the Cu or Ni sites directly facing the interfacial B or O atoms in the corresponding stacking configuration. The metal interlayer spacings are indicated in the side views. Red, green, blue, and gray spheres represent O, B, Cu, and Ni atoms, respectively.} 
\end{figure*}
\clearpage

\section{References}
\label{sec:references}

\renewcommand{\bibsection}{}

\bibliographystyle{apsrev4-1}  
\bibliography{SI}